\documentclass[twocolumn,superscriptaddress]{revtex4}

\usepackage{amsthm}

\usepackage{stix}

 \usepackage{amsfonts}
\usepackage{amsmath}
\usepackage{amssymb}
\usepackage{graphicx} 
\usepackage{dcolumn}
\usepackage{dsfont}
\usepackage{times} 
\usepackage{epsfig}
\usepackage[]{units}

\usepackage{lipsum}
 
\usepackage[colorlinks,citecolor=red,urlcolor=blue,bookmarks=false,hypertexnames=true]{hyperref}

\usepackage{color}
\definecolor{colorRed}{rgb}{0.75,0.,.0}
\definecolor{colorGreen}{rgb}{0.,0.75,.0}
\definecolor{colorBlue}{rgb}{0.,0,1}

\newcommand{\R}{{\mathbb R}}

\newcommand{\Z}{{\mathbb Z}}

\newcommand{\pair}[2]{\left(#1,#2\right)}
\newcommand{\tripple}[3]{\left(#1,#2,#3\right)}
\newcommand{\quadruple}[4]{\left(#1,#2,#3,#4\right)}

\def\sech{\mathop{\mathrm{sech}}}

\newcommand{\eps}{\varepsilon}

\def\epsilon{\varepsilon}

\def\beq{\begin{equation}}
\def\eeq{\end{equation}}

\def\calF{\mathcal{F}}
\def\calV{\mathcal{V}}
\def\calS{\mathcal{S}}
\def\calE{\mathcal{E}}
\def\calW{\mathcal{W}}
\def\calL{\mathcal{L}}

\def\sn{\mathrm{sn}}

\begin{document}

\title{Dispersive Shock Waves in Lattices: A Dimension Reduction Approach}

\author{Christopher Chong}
\affiliation{Department of Mathematics, Bowdoin College, Brunswick, Maine 04011}

\author{Michael Herrmann}
\affiliation{Institute of Partial Differential Equations, Technische Universit\"at Braunschweig, 38106 Braunschweig, Germany}

\author{P. G. Kevrekidis}
\affiliation{Department of Mathematics and Statistics, University of Massachusetts Amherst, Amherst, Massachusetts 01003-4515}

\begin{abstract}

Dispersive shock waves (DSWs), which connect states of different amplitude via a modulated wave train,  
form generically in nonlinear dispersive media subjected to
abrupt changes in state. 
The primary tool for the analytical study of DSWs is Whitham's modulation theory.
While this framework has been successfully employed in many space-continuous settings
to describe DSWs, the Whitham modulation equations are virtually intractable in most
spatially discrete systems.
In this article, 
we illustrate the relevance of
the reduction
of the DSW dynamics to a planar ODE 
in a broad class of lattice examples.
Solutions of this low-dimensional ODE 
accurately describe the orbits of the DSW in self-similar coordinates and the local 
averages in a manner consistent with the modulation
equations.
We use data-driven and quasi-continuum approaches within the context of a
discrete system of conservation laws to demonstrate how the underlying
low dimensional structure of DSWs can be identified and analyzed. 
The connection of these results to Whitham modulation theory is also discussed. 
\end{abstract}

\date{\today}

\vspace{1cm}

\maketitle

\section{Introduction}

Understanding how systems respond to sudden changes in the medium is critical in a number of applications.
Classic examples include explosions in shock tubes \cite{tube}, when gases or fluids are compressed
in a piston chamber \cite{piston} or when water dams break \cite{dam}.  From a modeling perspective,
a sudden change in a medium is often represented by step initial data (the so-called Riemann problem).
In mathematical models describing fluid or gas dynamics based on space and time continuous conservation laws, states of different amplitude that are connected via a discontinuity (i.e., fronts 
involving jumps, e.g., in density or concentration)
 are predicted to travel through the medium at constant speed and are called Lax shock waves \cite{Smoller}, or just shocks. 
Since solutions with infinite derivatives are nonphysical, modified (e.g., regularized) models are often used as an alternative.
In some settings, such as stratified environments of the ocean and atmosphere, dispersive shocks (and their regularizations) are more relevant \cite{Hoefer2016}. 
 In this case, the states of different amplitude are connected 
via an expanding modulated wave train. This structure is called a dispersive shock wave (DSW).
The primary tool to analytically describe DSWs is Whitham's modulation theory \cite{Whitham74,GP73,Karpman}.
In this framework, one derives equations describing slow modulations of the underlying parameters of a periodic wave by, for example, averaging 
the Lagragian action integral over a family of periodic wave trains \cite{dsw,Mark2016}. 

The study of DSWs in spatially continuous media has been ongoing since Whitham's seminal work \cite{Whitham74} over 50 years ago,
but there has been renewed excitement for this theme. This excitment has largely been inspired by groundbreaking experimental observations of DSWs 
most notably in ultracold gases and superfluids~\cite{Hoefer2006,davis}, 
nonlinear optics~\cite{fleischer,Trillo2018}, and fluid 
conduits~\cite{Hoefer2016,Hoefer2018}, among others.
Dispersive shock waves in one-dimensional (1D) nonlinear lattices (to be called lattice DSWs) have been explored numerically, and even experimentally in several works \cite{first_DSW,Nester2001,Hascoet2000,Herbold07,Molinari2009,shock_trans_granular,HEC_DSW}. 
Although much of the above motivation stems from the material
science of granular
crystals, it is of broad physical interest, as similar
structures have been experimentally observed, e.g., in nonlinear
optics of waveguide arrays~\cite{fleischer2}.
The existence of periodic waves has been proved \cite{Iooss2000,Pankov05,Herrmann10b} and corresponding 
 modulation equations have been derived \cite{Venakides99,DHM06}.
Explicit forms of the periodic waves are typically not available, resulting in modulation equations that 
are virtually intractable. Even in the integrable 
paradigm of the Toda lattice,
the modulation equations are quite cumbersome \cite{Bloch_Toda92,Venakides91,Holian81}.
This is one reason the mathematical theory of lattice DSWs is not as mature when compared to
the study of DSWs in continuous settings, which has seen many recent experimental and theoretical
advancements, as summarized in the reviews \cite{scholar,Mark2016,dsw}. 

A first goal of this paper is to provide a complementary approach to modulation theory for the description of lattice DSWs
in a broad class of
nonlinear lattice dynamical models. Subsequently, we aim
to unveil the crucial observation that
lattice DSW dynamics can be reduced 
to a planar ODE which can be handled analytically. 
After detailing the problem set-up in Sec.~\ref{sec:model}, we explore
two approaches towards identifying the underlying ODE dynamics: a data driven one in Sec.~\ref{sec:datafit} and one based
on a quasi-continuum approximation in Sec.~\ref{sec:quasi}. The obtained results are discussed 
within the context of the modulation equations in Sec.~\ref{sec:modulation}. Section~\ref{sec:theend} concludes the paper.
We believe that the method presented herein offers a key insight
of relevance to a wide class of lattice models bearing DSWs
and the method of analysis offers a quantitative perspective on
the DSW dynamics that is found to be in  good agreement
with the numerical observations in suitable regimes of the
wave speed $c$. A discussion of the modulation
equations in terms of the wave parameters is discussed
in further detail in Appendix \ref{app:mod}.

 \section{Model Equations and Problem Statement} \label{sec:model}

In order to explore lattice DSWs and the dimensionality reduction approach, 
 the following wide class of nonlinear dynamical
 lattices is considered~\cite{wilma}
\begin{equation} \label{deq}
2 \frac{du_n}{dt} +  \Phi'(u_{n+1} )-  \Phi'(u_{n-1} ) =0,
\end{equation}
where $n\in \Z,  t\in \R, u=u_n(t) \in \R$ and the potential $\Phi(u)$ is assumed to be convex. Equation~\eqref{deq} possesses a Lagragian and Hamiltonian structure \cite{Herrmann_Scalar}, yet it is first
order only, making its analysis slightly more convenient when compared to classical
 nonlinear oscillators, such as those of the Fermi-Pasta-Ulam-Tsignou (FPUT) type \cite{FPU55}. We expect that the reduction approach that is presented will be applicable to a large class of lattice systems, such as FPUT models \cite{FPUreview}, discrete Nonlinear Schr\"odinger models \cite{PK09}, and
 discrete Klein Gordon models \cite{floyd}. Besides serving as a prototype
 model for lattice DSWs, Eq.~\eqref{deq} is also directly relevant for applications, such as in the description of traffic flow \cite{Whitham90}; for a discussion of relevant models and their
 continuum limits see also Ref.~\cite{wilma}.
Equation~\eqref{deq} is a centered difference scheme for the space and time continuous conservation law
\begin{equation}\label{eq}
\partial_T U + \partial_X \Phi' ( U ) = 0,\end{equation}
where  $X,T  \in \R$, $U=U(X,T) \in \R$, and
the discrete and continuous variable are related through $u_n(t) = U(\epsilon n, \epsilon t) = U(X, T)$ with $0<\epsilon \ll 1$.
It is well known that Lax shocks can form in Eq.~\eqref{eq} subjected to Riemann initial data for a large class of flux functions
$\Phi'$ \cite{Smoller}. Equation~\eqref{deq} can be thought of as a dispersive regularization of  Eq.~\eqref{eq} \cite{LAX86},
and hence, it is expected that dispersive shock waves will form when subjected to Riemann initial data instead of Lax shocks~\cite{wilma}.

For demonstration purposes, we will primarily consider the polynomial potential
\begin{equation}\label{pot}
\Phi(u) =\frac{u^2}{2} + \frac{u^4}{4},
\end{equation}
although, we also present results for the Kac-van-Moerbeke (KvM) potential,
$\Phi(u) = \exp(u)$, to demonstrate the generality of the presented approach.
Note that Eq.~\eqref{deq} with the KvM potential is completely integrable \cite{KAC1975160,Whitham90} although this fact
will play no role in the analysis, as the methods
presented below are of interest more broadly to 
dispersive nonlinear lattice systems (rather than
more restrictively integrable ones).
 \begin{figure}[t]
    \centering
   \begin{tabular}{@{}p{0.5\linewidth}@{}p{0.5\linewidth}@{}  }
  \rlap{\hspace*{5pt}\raisebox{\dimexpr\ht1-.1\baselineskip}{\bf (a)}}
 \includegraphics[height=3.3cm]{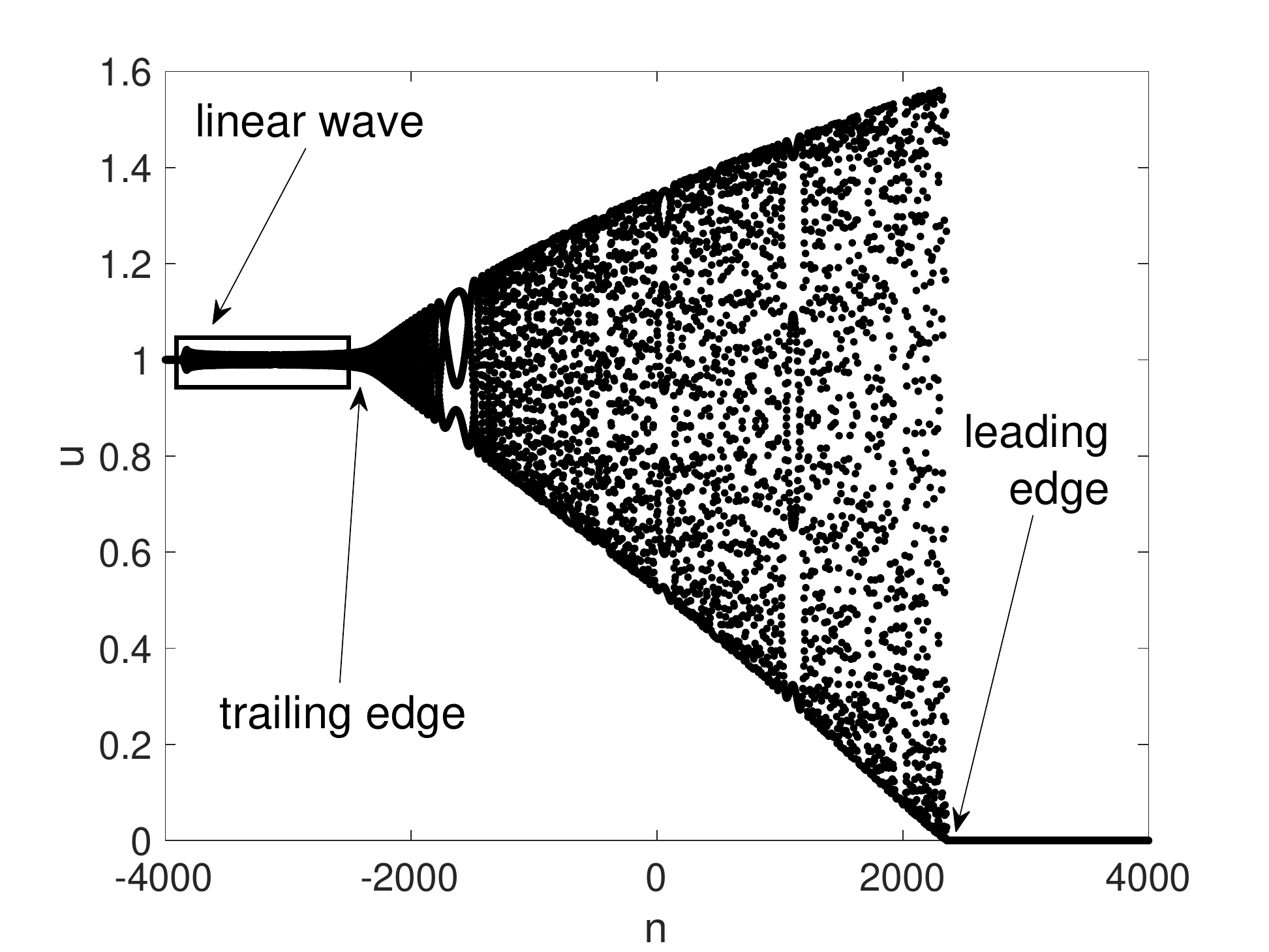} &
  \rlap{\hspace*{5pt}\raisebox{\dimexpr\ht1-.1\baselineskip}{\bf (b)}}
\includegraphics[height=3.3cm]{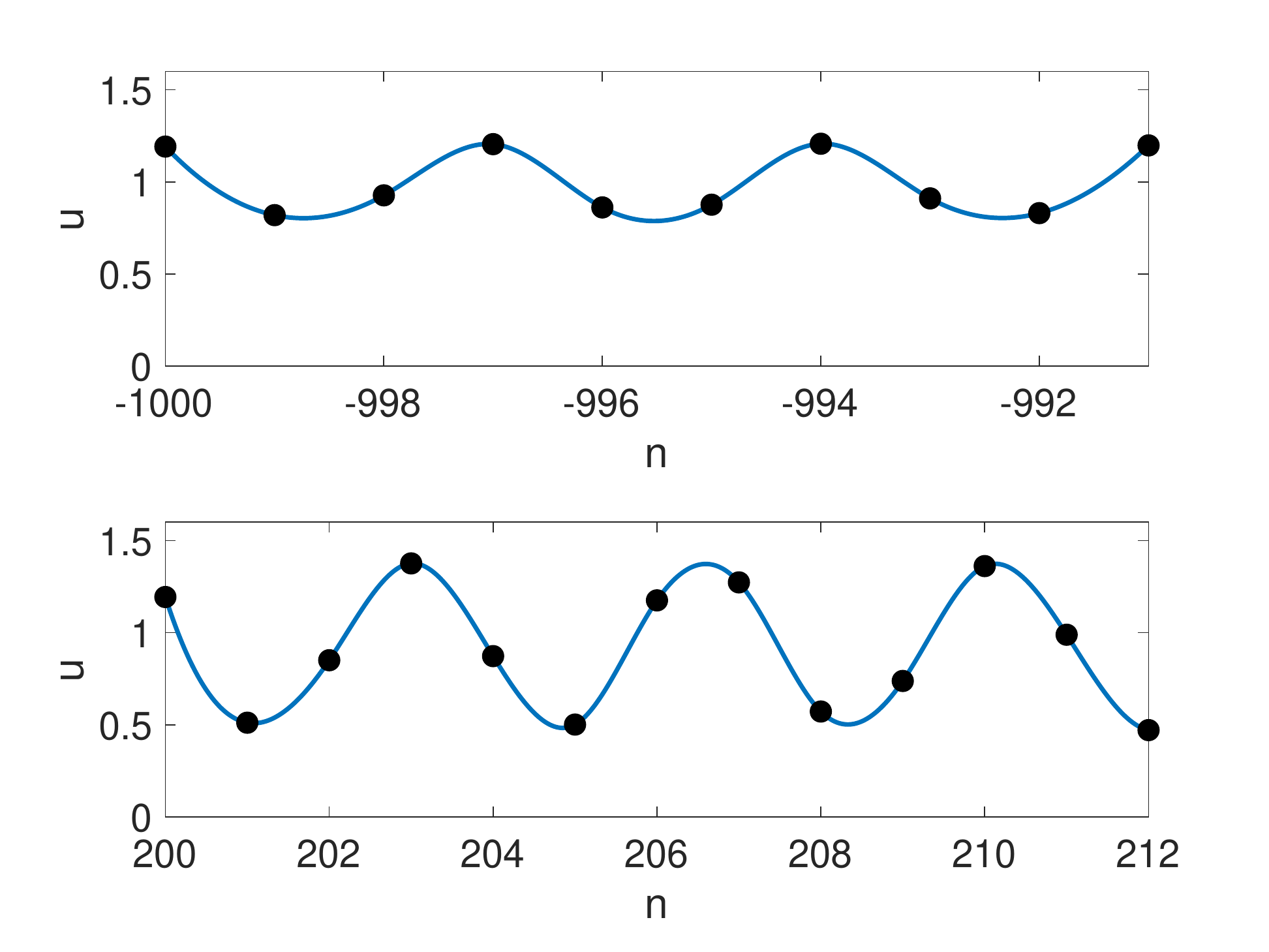}
  \end{tabular}
 \caption{\textbf{(a)} Example of lattice DSW. Result obtained via numerical simulation of Eq.~\eqref{deq} with $N=8000$ and the polynomial potential, Eq.~\eqref{pot}, starting from Riemann initial data, Eq.~\eqref{step}. Solution at $t=960$ shown.  
 The box labeled ``linear wave" vanishes in the large lattice limit. \textbf{(b)} Zoom of panel (a) at near $n=-1000$ (top) and $n=200$ (bottom). The solid lines are the interpolating splines, shown for visual clarity.
The profiles resemble periodic waves, albeit with different parameters, in each zoom.
 }
 \label{fig:dsws}
\end{figure} 
  \begin{figure*}[t]
\begin{tabular}{@{}p{0.33\linewidth}@{}p{0.33\linewidth}@{}p{0.33\linewidth}@{}  }
  \rlap{\hspace*{5pt}\raisebox{\dimexpr\ht1-.1\baselineskip}{\bf (a)}}
 \includegraphics[height=4.5cm]{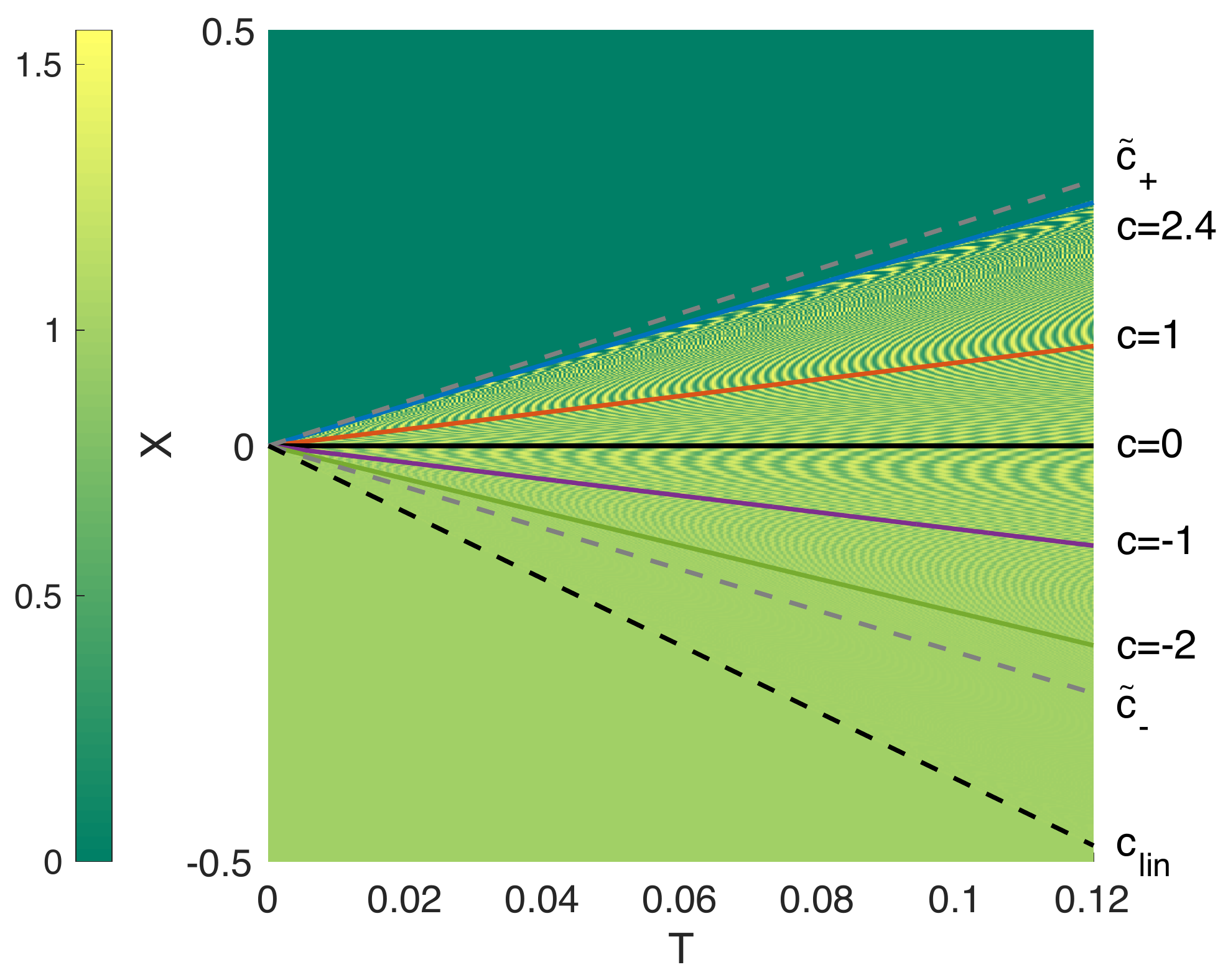} &
  \rlap{\hspace*{5pt}\raisebox{\dimexpr\ht1-.1\baselineskip}{\bf (b)}}
\includegraphics[height=4.5cm]{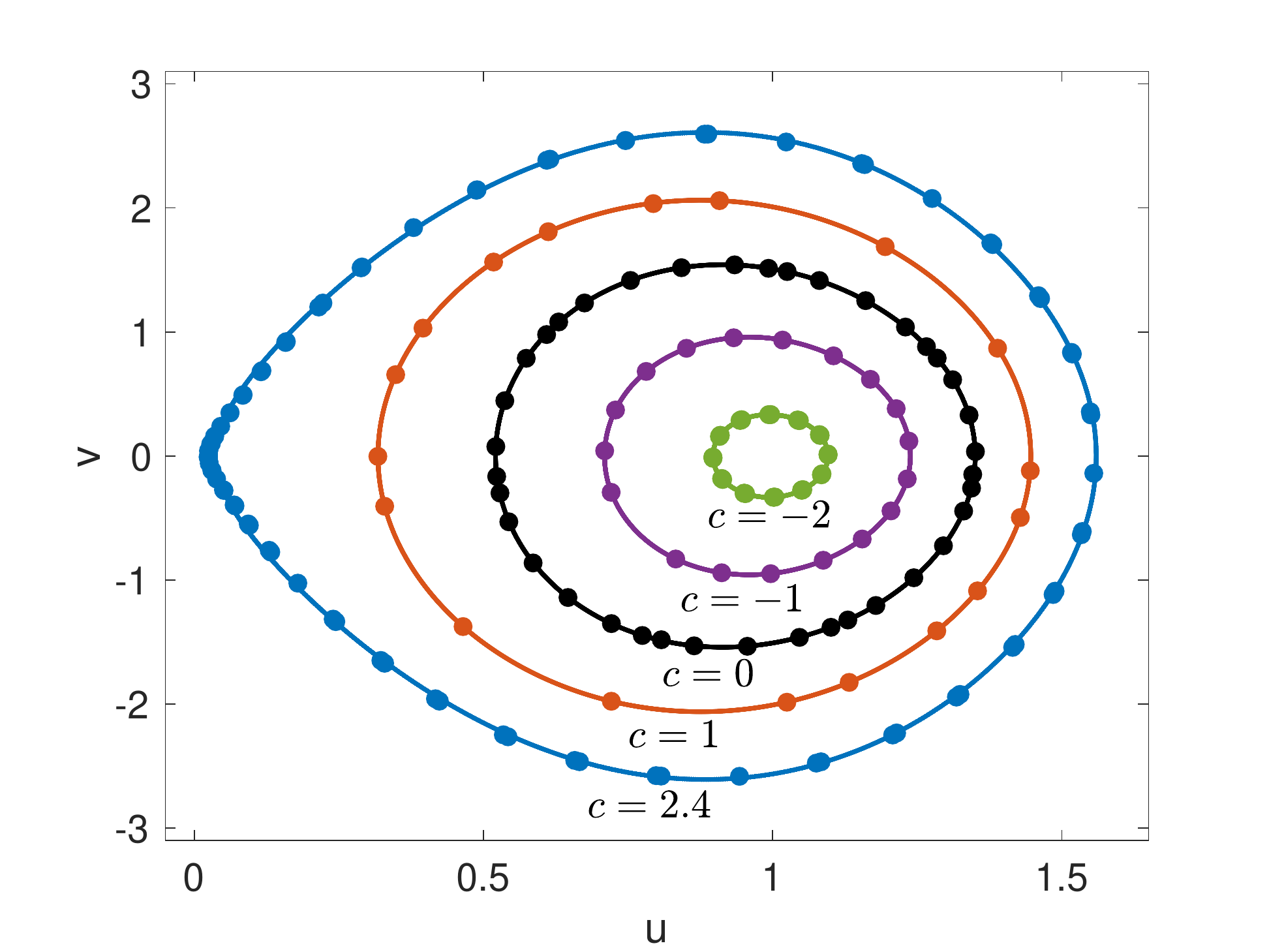}&
  \rlap{\hspace*{5pt}\raisebox{\dimexpr\ht1-.1\baselineskip}{\bf (c)}}
\includegraphics[height=4.5cm]{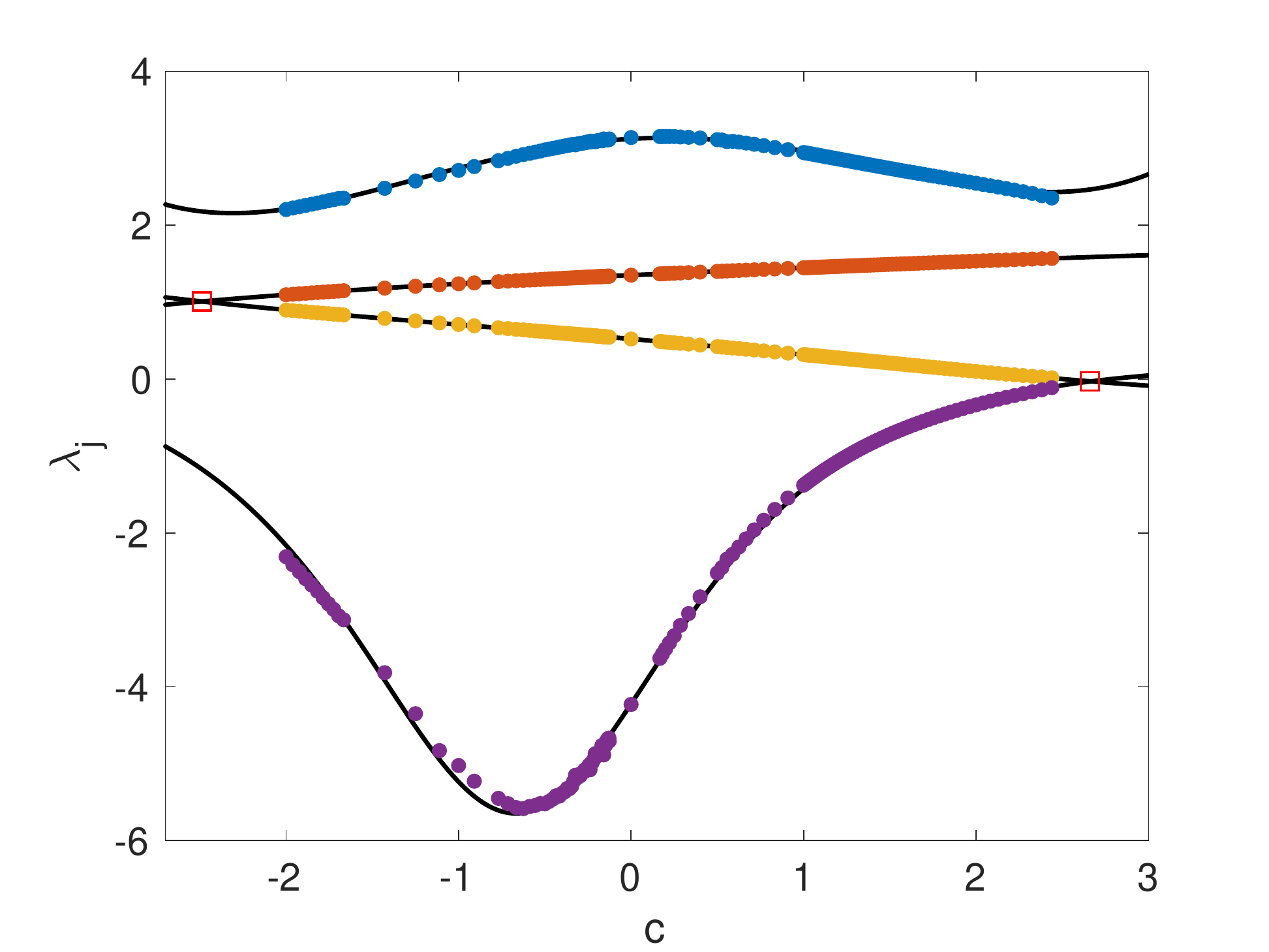}
  \end{tabular}
 \caption{\textbf{(a)} Intensity plot of a lattice DSW (same parameters as in Fig.~\ref{fig:dsws}) with axes corresponding to the slow variables $X=\epsilon n$ and $T = \epsilon t$. Color intensity corresponds to amplitude (see colorbar). Each solid line corresponds to a fixed value of the self-similar speed $c = X/T$. The dashed lines marked $\tilde{c}_+$ and $\tilde{c}_-$ are estimates of the leading and trailing edge of the DSW, respectively. The dashed line marked $c_{\rm lin}$ is the edge of the linear wave (see boxed area of Fig.~\ref{fig:dsws}(a)). The vertical slice at the final time shown ($T=0.12$) corresponds to the profile in Fig.~\ref{fig:dsws}(a). \textbf{(b)}  Phase plane of the data extracted along the lines $c=X/T$ of the lattice DSW shown in panel (a) (markers),
 where $v = \dot{u}$. The solid lines are the planar ODE prediction of Eq.~\eqref{odesol}.
 \textbf{(c)} Plot of the roots (markers) of $P(c)$. 
 The roots are ordered by size: $\lambda_1$ (purple) $ \leq \lambda_2$ (yellow) $\leq \lambda_3$ (red) $\leq \lambda_4$ (blue). Solid lines are the result
 of fitting the roots to a trial function of $c$; see
 relevant details in the text.
  }
 \label{fig:cTfix_data}
\end{figure*} 
Throughout the manuscript we consider Riemann (i.e. step) initial data,
\begin{equation}\label{step}
u_n(0) = \left\{
\begin{split}
1, \quad & n \leq 0 \\
0, \quad & n > 0
\end{split}
\right.
\end{equation}
to generate DSWs. Letting $N$ (even) represent the number of spatial points, we define $\epsilon = 1/N$. The corresponding spatial domain is
$-N/2+1 < n \leq N/2$ and the temporal domain is $[0, T_f / \eps]$, where $T_f$ is a fixed constant independent of $\epsilon$. 
One example of
a DSW that has formed starting with \eqref{step} is shown in Fig.~\ref{fig:dsws}(a). For every finite lattice,
there is a linear wave centered at unity (see boxed area in Fig.~\ref{fig:dsws}(a)), which vanishes
according to the decay law $\sim N^{-1/3}$, see Appendix \ref{app:tail}. The trajectory of the DSW in each small window of space and time appears to be a traveling periodic wave, see Fig.~\ref{fig:dsws}(b). The existence of traveling periodic waves of Eq.~\eqref{deq} was
established in Ref.~\cite{Herrmann_Scalar}. They have the form
\begin{equation}\label{tw}
u_{n}(t)=V+\mathcal{V}(\zeta), \qquad \zeta = K n-  \Omega t,
\end{equation}
where the parameters $V$ (mean), $K$ (wave vector) and $  \Omega$ (frequency) are real and
 the wave profile $\mathcal{V}$ is assumed, w.l.o.g, to be $2\pi$ periodic with mean zero. 
 By substituting Eq.~\eqref{tw} into Eq.~\eqref{deq}, one finds that $\mathcal{V}$ must satisfy the advance-delay equation,
\begin{equation} \label{adv1}
2 \Omega \mathcal{V}^{\prime}(\zeta)=\Phi^{\prime}(V+\mathcal{V}(\zeta+K))-\Phi^{\prime}(V+\mathcal{V}(\zeta-K)).
\end{equation}
 Within the ``core" of the DSW (i.e., the modulated wave connecting the two constant states), the wave parameters, $V$, $K$ and $  \Omega$ appear to be different in each small spatiotemporal window
(compare the top and bottom panels of Fig.~\ref{fig:dsws}(b)). This suggests that a DSW can be viewed as a traveling periodic wave with slowly varying parameters that connect the left state $u_- = 1$ to the right state $u_+ = 0$.
 The set of equations that describe how the wave parameters, $V$, $K$ and $  \Omega$ vary within
 the core of the DSW is referred to as the system of modulation equations (relevant details are given in Sec.~\ref{sec:modulation} and Appendix \ref{app:mod}).
There is no explicit form for the traveling wave profile $\mathcal{V}$ for general
potentials $\Phi(u)$, which satisfies the above advance-delay differential equation~(\ref{adv1}).
This results in modulation equations that are intractable in the discrete setting.
Thus, in order to obtain a useful analytical description of a lattice DSW, an approach that is complementary to the modulation theory seems necessary. We now discuss such an approach.

\section{A Planar ODE description of a lattice DSW} \label{sec:datafit}

Figure~\ref{fig:cTfix_data}(a) shows an intensity plot of a numerical solution of Eq.~\eqref{deq} with initial data given by
Eq.~\eqref{step} and $\Phi(\phi) = \frac{\phi^2}{2} + \frac{\phi^4}{4}$ and $N=8000$. A symplectic integrator 
is used for the direct simulation of Eq.~\eqref{deq}, see \cite{Herrmann_Scalar}.  Viewing the solution along slices of fixed $t$ yields the usual DSW spatial profile, see Fig.~\ref{fig:dsws}(a). However, if one observes data along the DSW with $n/t = X/T= c$ fixed,
a nested, non-intersecting set of closed loops in the phase plane $(u,\dot{u})$ is formed, with each value of $c$ representing a different loop,
see Fig~\ref{fig:cTfix_data}(b). The fact that closed loops form is unsurprising, since it is expected that the DSW has a self similar
structure, and hence, with $X/T = c$ fixed, the modulation parameters should be fixed. What is surprising is that the set of loops
is nested and non-intersecting when projecting the dynamics onto a two-dimensional phase space. This motivates
the proposal to identify a planar ODE that can approximate the lattice DSW dynamics.

To identify such an ODE, we first take a data-driven approach.
For each fixed value of $c$, we extract position and velocity data from the DSW. we define this data
as $u = u_{c t}(t ;c)$, where $u_{c t}(t ;c)$ is the position data at lattice location $n=c t$ and time $t$, which
ensures that $c = n/t$. This necessitates that the time increment is chosen such that $c t$ is an integer. 
Similarly, we define $v = v_{c t}(t ;c)$, where $v_n = \dot{u}_n$. 
 We assume there is a simple energy relationship 
between the position $u$ and velocity $v$ as follows
\begin{equation} \label{energy1}
\frac{v^2}{2} = 
 \sum_{j=0}^M \nu_j u^j(t;c) =: P(u;c).
\end{equation}
Treating $u$ and $v$ as independent and dependent variables,
respectively, we determine
the parameters $\nu_j$ via a linear regression. This is repeated for several $c$ values
that lie within the core of the DSW.  For each value of $c$,  we express the polynomial $P$
in terms of its roots
$P(u;c) = \Lambda(c) (u - \lambda_1(c))(u - \lambda_2(c))(u - \lambda_3(c))(u - \lambda_4(c))$
where the roots are ordered $\lambda_1 \leq \lambda_ 2 \leq \lambda_3 \leq \lambda_4$.
The roots plotted against the speed $c$ are the markers shown in Fig.~\ref{fig:cTfix_data}(c). 
Once the roots and constant 
(i.e., $u$-independent)
factor $\Lambda$ are obtained, they are fitted to a simple trial function of $c$. 
For the purpose of simplicity, polynomial trial functions are assumed, which 
capture the data well (e.g., compare the lines and markers of the top three curves of Fig.~\ref{fig:cTfix_data}(c)). 
The only exception of this choice of trial function was for the first root, $\lambda_1$, in the case
of the polynomial potential. This root is bell-shaped (see bottom curve Fig.~\ref{fig:cTfix_data}(c)),
and thus, the polynomial trial function led to a poor fit. Thus, a trial 
function in the form of a  hyperbolic secant is used in this case. All 
other cases led to good agreement between the roots and a polynomial
trial function. Note that the choice of trial function has no direct
consequence on the forthcoming analysis. Upon performing the fitting, the resulting
formulas are  $\lambda_1(c)  =    -5.79 \sech(  -1.16 ( c + 0.67 )) +   0.20 $, 
$\lambda_2(c) =  -0.01c^2   - 0.20 c +    0.52$, $\lambda_3(c) =  -0.01 c^2   +  0.10 c + 1.35$, 
$\lambda_4(c) =    0.02 c^3 - 0.19c^2   +   0.05 c +    3.08$ and 
$\Lambda(c) =  0.40 c^2   + 0.42c   + 0.62$. Repeating the entire procedure for the KvM potential   
$\Phi(u) = \exp(u)$ leads to the formulas $\lambda_1(c)  =  0.03 c^3 - 0.42c^2   +   1.74 c -   2.08$
(notice that in this case, a polynomial was used
instead), 
$\lambda_2(c) =  0.01 c^3 - 0.01c^2   -   0.38 c +  0.72$, 
$\lambda_3(c) =   -0.01 c^3 - 0.02c^2   +  0.29 c +   1.24$, 
$\lambda_4(c) =    0.04 c^3 - 0.23c^2   +   0.06 c +    3.43$ 
and $\Lambda(c) =  0.13 c^2   + 0.25c   + 0.49$.

  \begin{figure*}[t]
    \centering
     \begin{tabular}{@{}p{0.5\linewidth}@{}p{0.5\linewidth}@{}  }
  \rlap{\hspace*{5pt}\raisebox{\dimexpr\ht1-.1\baselineskip}{\bf (a)}}
 \includegraphics[height=6cm]{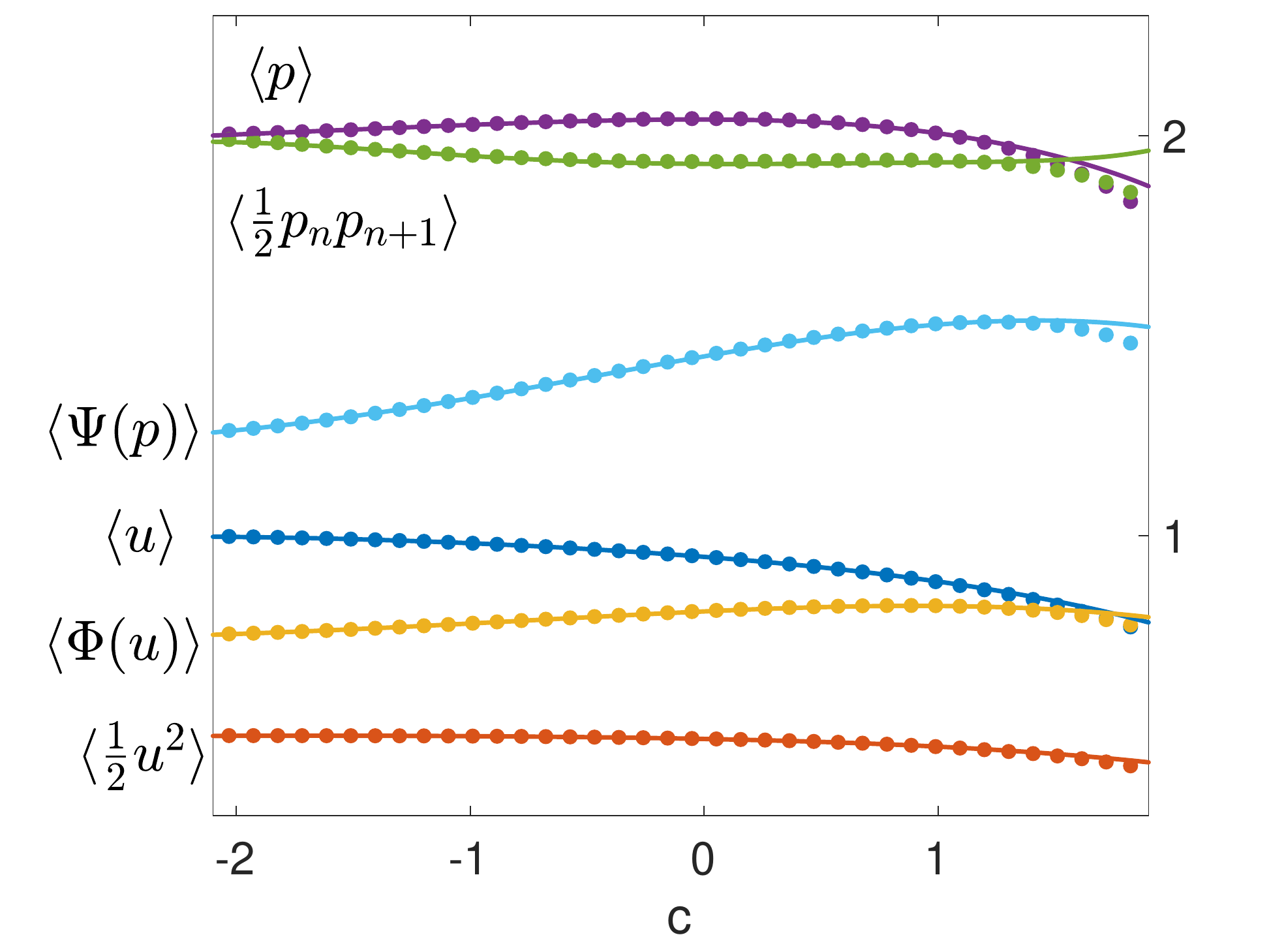} &
  \rlap{\hspace*{5pt}\raisebox{\dimexpr\ht1-.1\baselineskip}{\bf (b)}}
\includegraphics[height=6cm]{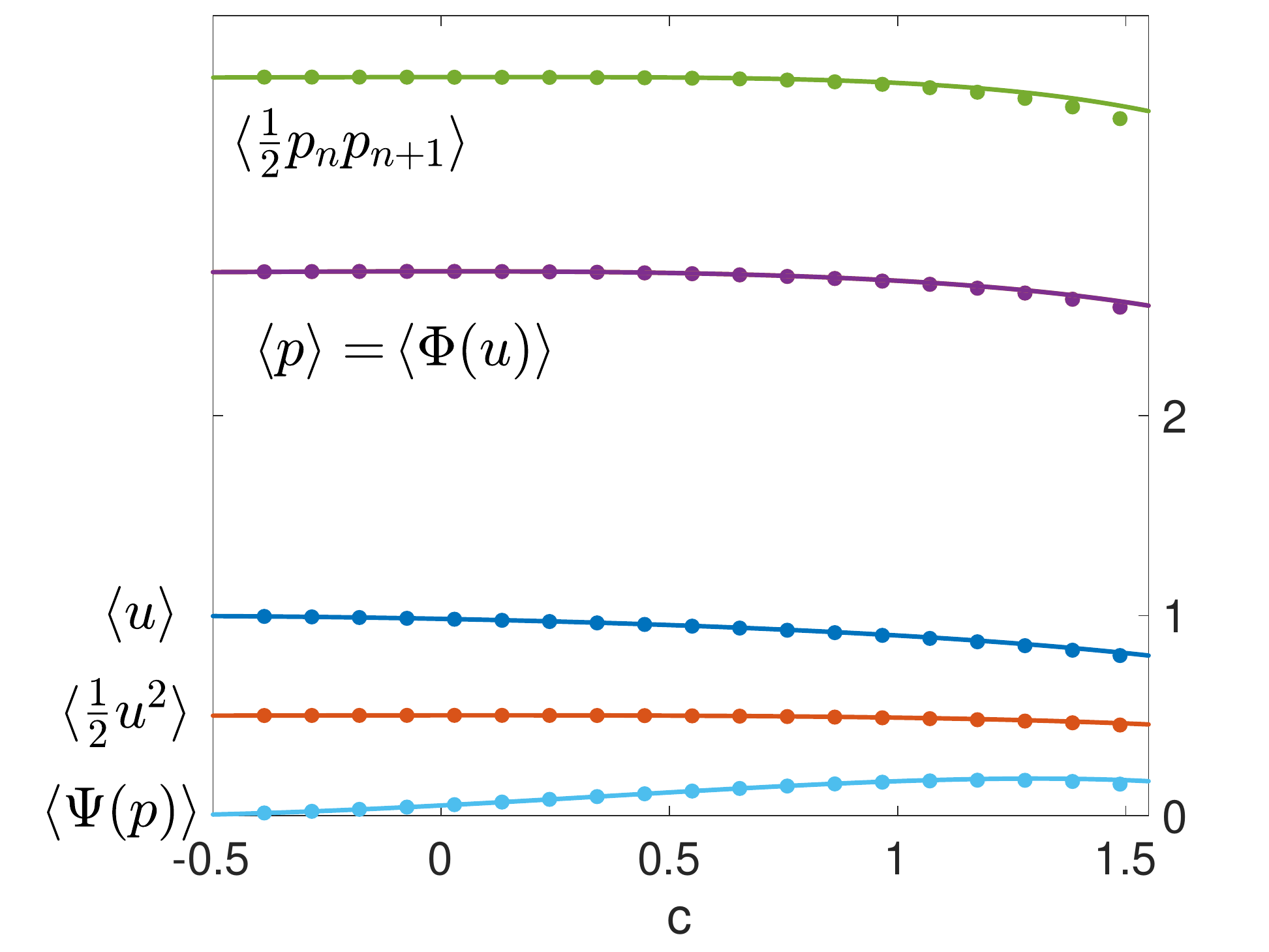}
  \end{tabular}
 \caption{
 \textbf{(a)} Local averages measured from the lattice DSW (markers) and
 the ODE prediction from Eq.~\eqref{odesol} (solid curves)
  for the polynomial potential $\Phi(u) = u^2/2 + u^4/4$.
\textbf{(b)}:  Same as panel (a) for the KvM potential $\Phi(u) = \exp(u)$.
}
 \label{fig:local}
\end{figure*} 

Equation~\eqref{energy1} implies the following planar ODE 
is satisfied
\begin{equation} \label{lowdim}
\frac{d^2u}{d \tau^2}  =  P'(u;c), \quad u(0) = \lambda_2(c), \quad \left. \frac{du}{d\tau} \right | _{\tau = 0} = 0 , 
\end{equation}
with $\tau \in[0,\tau_p]$ where $\tau_p$ is the period of oscillation and $\tau$ is related to the lattice time variable
through the scaling $\tau = \alpha t$. While the time scaling factor is not needed for the forthcoming analysis, it
can be approximated through the relation $\alpha = \tau_p/(T_p)$, where $T_p$ is the period of oscillation
of the DSW trajectory with $c$ fixed. Note that both $\tau_p$ and $T_p$ (and hence $\alpha$) depend on $c$.
Clearly, periodic waves of Eq.~\eqref{lowdim} will correspond to oscillations of $u$
between roots of $P(u;c)$ when $P>0$, such as  $\lambda_2(c)$ and $\lambda_3(c)$. With the choice of initial values
 these extreme values occur for $t=0$ and $t=\tau_p/2$
such that $u(0) = \lambda_2$ is the minimum value and $u(\tau_p/2) = \lambda_3$ is the maximum.  The period of oscillation is given by
$ \tau_p =  2\int_{0}^{\tau_p/2}  d\tau =2 \int_{\lambda_2}^{\lambda_3} 1/ \sqrt{2 P(u)} \, du $. 
For the choice of the energy relationship given by Eq.~\eqref{energy1}, Eq.~\eqref{lowdim} can be solved using quadrature \cite{Kamchatnov}.
The exact solution is
\begin{equation} u(\tau) = \frac{\lambda_2 (\lambda_3 - \lambda_1) + \lambda_1 (\lambda_2 - \lambda_3) \sn^2(\theta \tau; m )  }{ \lambda_3 - \lambda_1 +(\lambda_2 - \lambda_3) \sn^2(\theta \tau; m )}, \label{odesol} 
\end{equation}
where $\sn$ is a Jacobi elliptic function with parameter $0<m<1$ and,
$$ \theta =  \sqrt{  \frac{ \Lambda(\lambda_3-\lambda_1)(\lambda_4-\lambda_2)}{2} }, \qquad m = \frac{ (\lambda_3-\lambda_2)(\lambda_4-\lambda_1)}{ (\lambda_3-\lambda_1)(\lambda_4-\lambda_2)}. $$
The periodic wave given by equation~\eqref{odesol} has the period $\tau_p = 2K(m)/\theta$, where $K(m)$ is the complete elliptic integral of the first kind. Equation~\eqref{odesol} yields good agreement
with the trajectories of the full lattice DSW dynamics, despite the simple choice of a polynomial for the energy relationship given by Eq.~\eqref{energy1}, (compare the lines and markers of Fig.~\ref{fig:cTfix_data}(b)).
Notice for $\lambda_1 \rightarrow \lambda_2$ we have $m \rightarrow 1$, and thus, the period tends to infinity and the solution tends to a homoclinic connection.
This limit corresponds to the leading edge of the DSW. For the polynomial potential, the ODE prediction is $\tilde{c}_+ = 2.66$,
which overestimates the observed value of  $c_+ = 2.45$.
 For $\lambda_3 \rightarrow \lambda_2$ we have $m \rightarrow 0$, in which case the solution tends to a  constant.
This is the trailing edge of the DSW, which is predicted by the ODE to occur for $\tilde{c}_- = -2.48$, which is
quite close to the value extracted from the full lattice DSW $c_- = -2.50$. See the red boxes of Fig.~\ref{fig:cTfix_data}(c) showing where the roots coalesce and the gray dashed lines of Fig.~\ref{fig:cTfix_data}(a) for a comparison of the predicted and actual DSW edges.
For the KvM potential, the trailing and leading edge extracted from the lattice DSW are $c_- = -0.71$  and $c_+ = 1.98$,
and the ODE predictions are $\tilde{c}_- = -0.73$ and $\tilde{c}_+ = 2.10$.

\subsection{ODE prediction of local averages}

  \begin{figure*}
 \begin{tabular}{@{}p{0.25\linewidth}@{}p{0.25\linewidth}@{}p{0.25\linewidth}@{}p{0.25\linewidth}@{}  }
  \rlap{\hspace*{5pt}\raisebox{\dimexpr\ht1-.1\baselineskip}{\bf (a)}}
 \includegraphics[height=3.5cm]{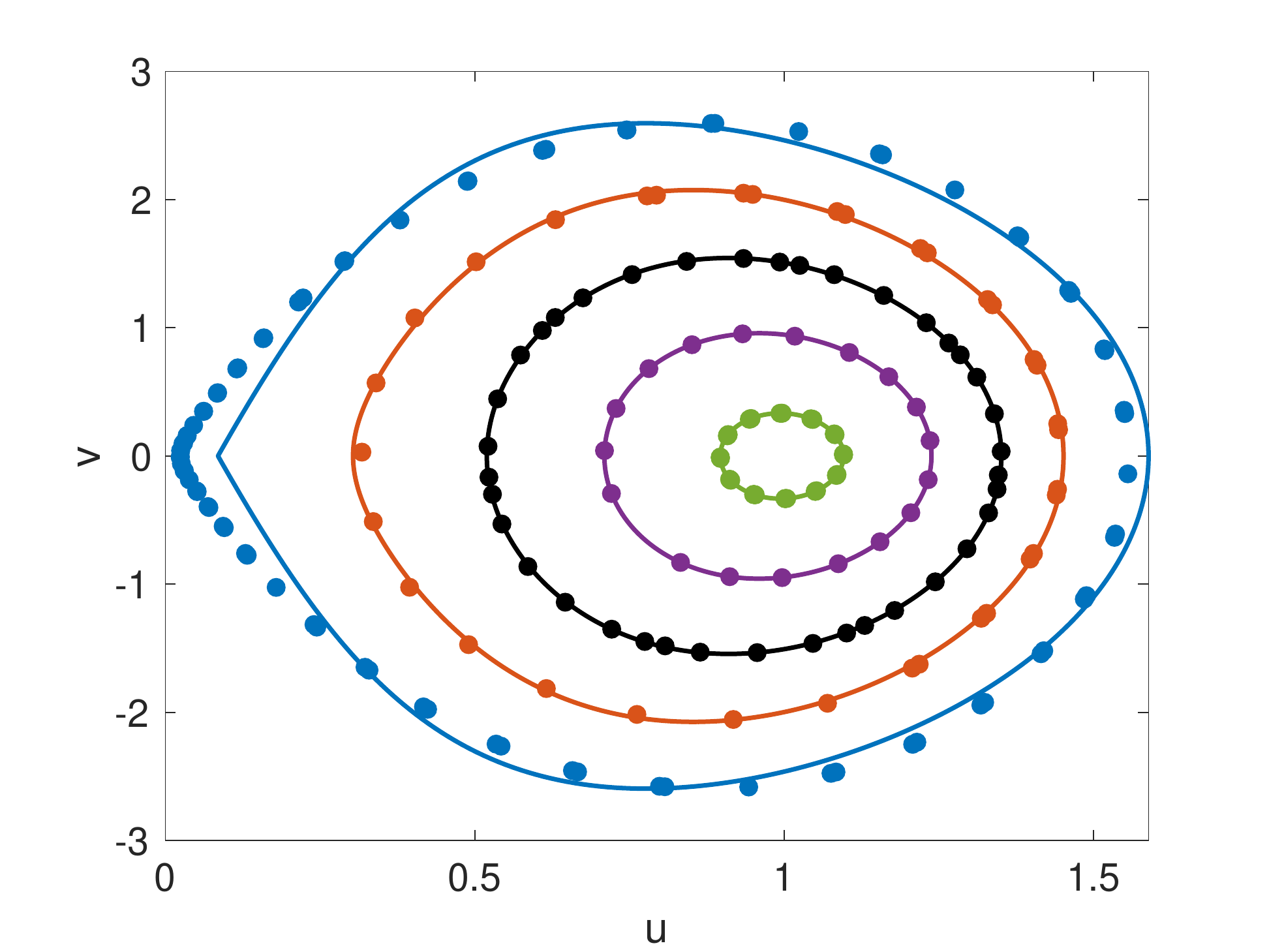} &
  \rlap{\hspace*{5pt}\raisebox{\dimexpr\ht1-.1\baselineskip}{\bf (b)}}
\includegraphics[height=3.5cm]{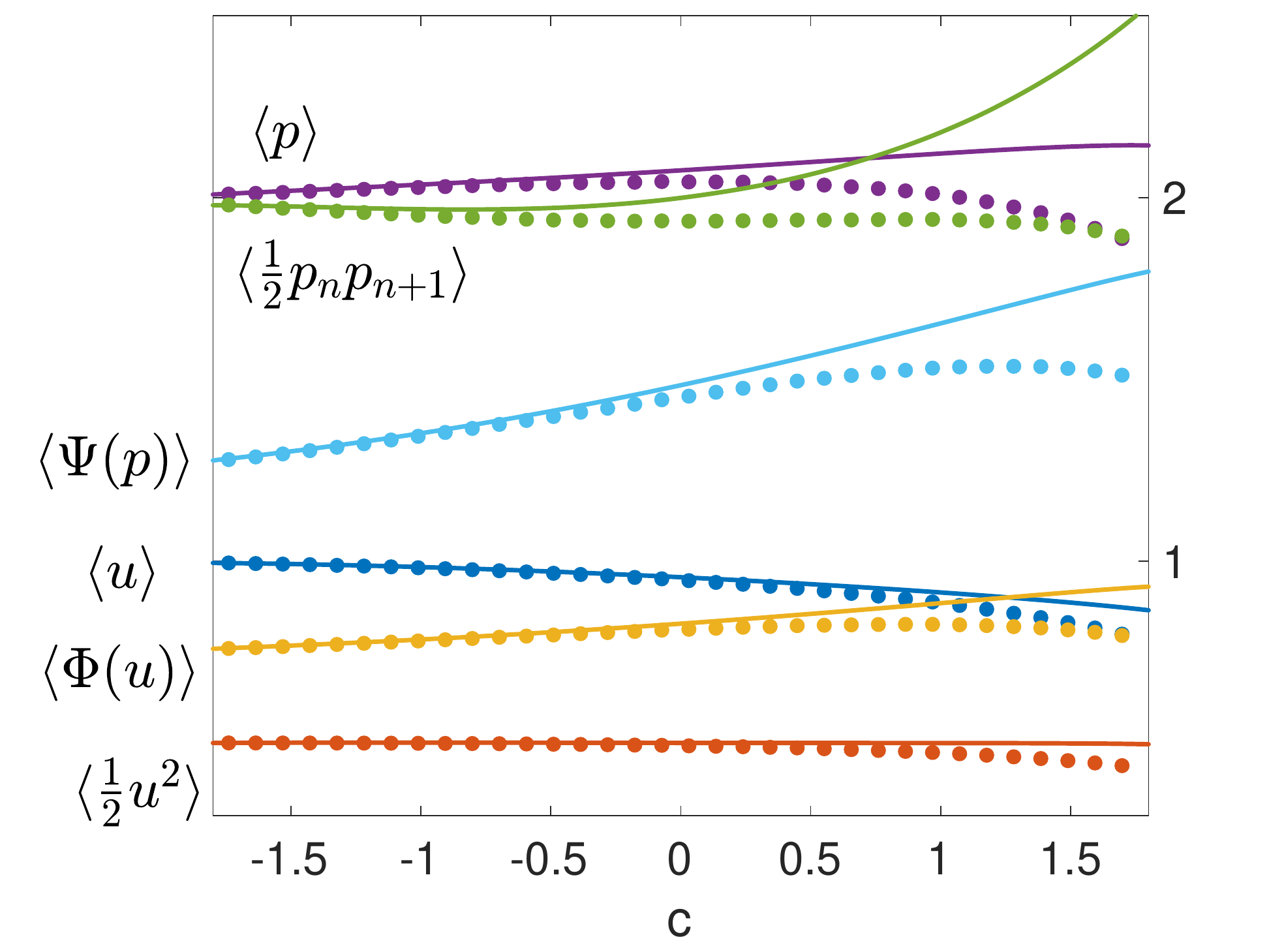}&
  \rlap{\hspace*{5pt}\raisebox{\dimexpr\ht1-.1\baselineskip}{\bf (c)}}
\includegraphics[height=3.5cm]{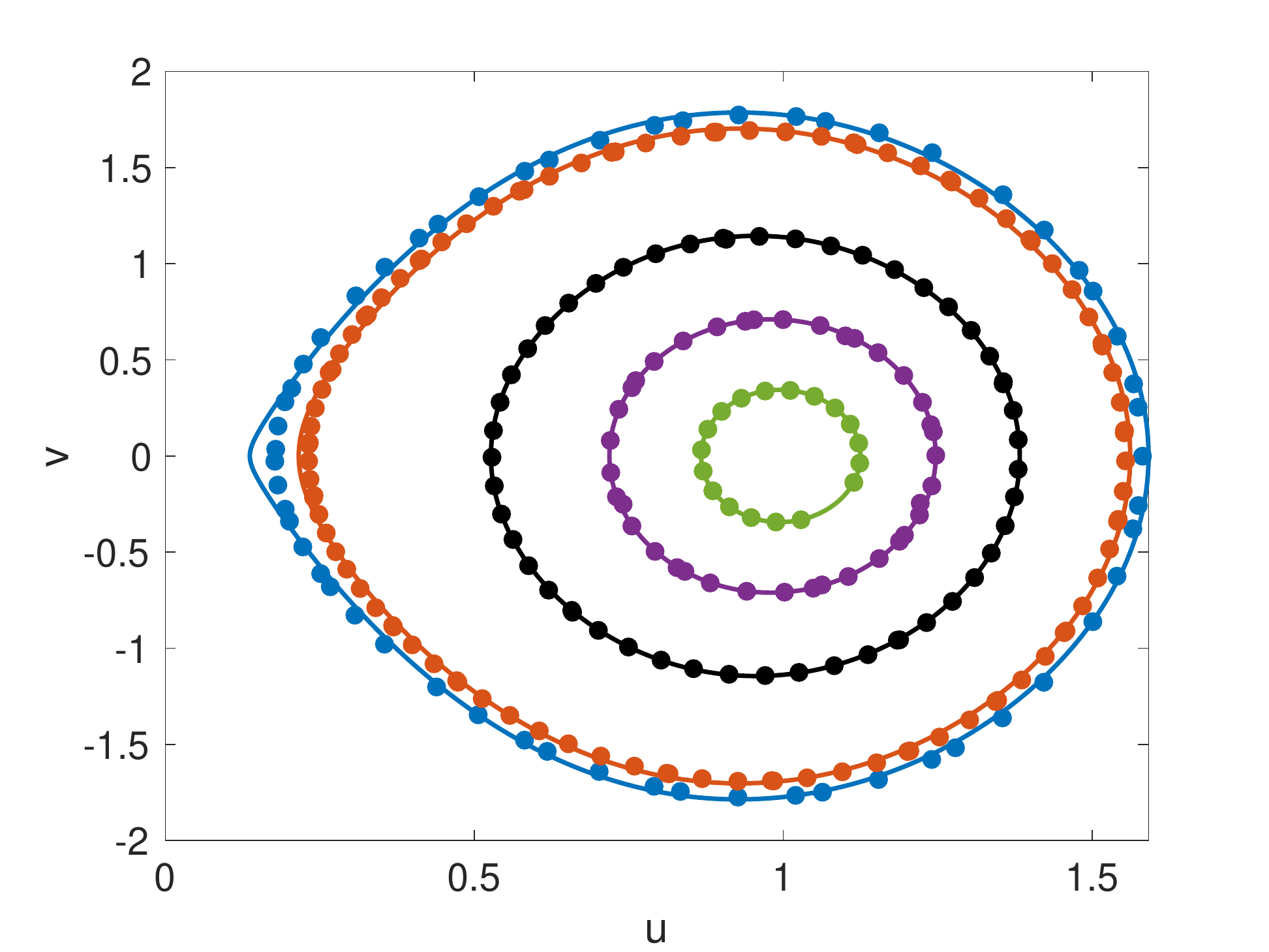} &
  \rlap{\hspace*{5pt}\raisebox{\dimexpr\ht1-.1\baselineskip}{\bf (d)}}
\includegraphics[height=3.5cm]{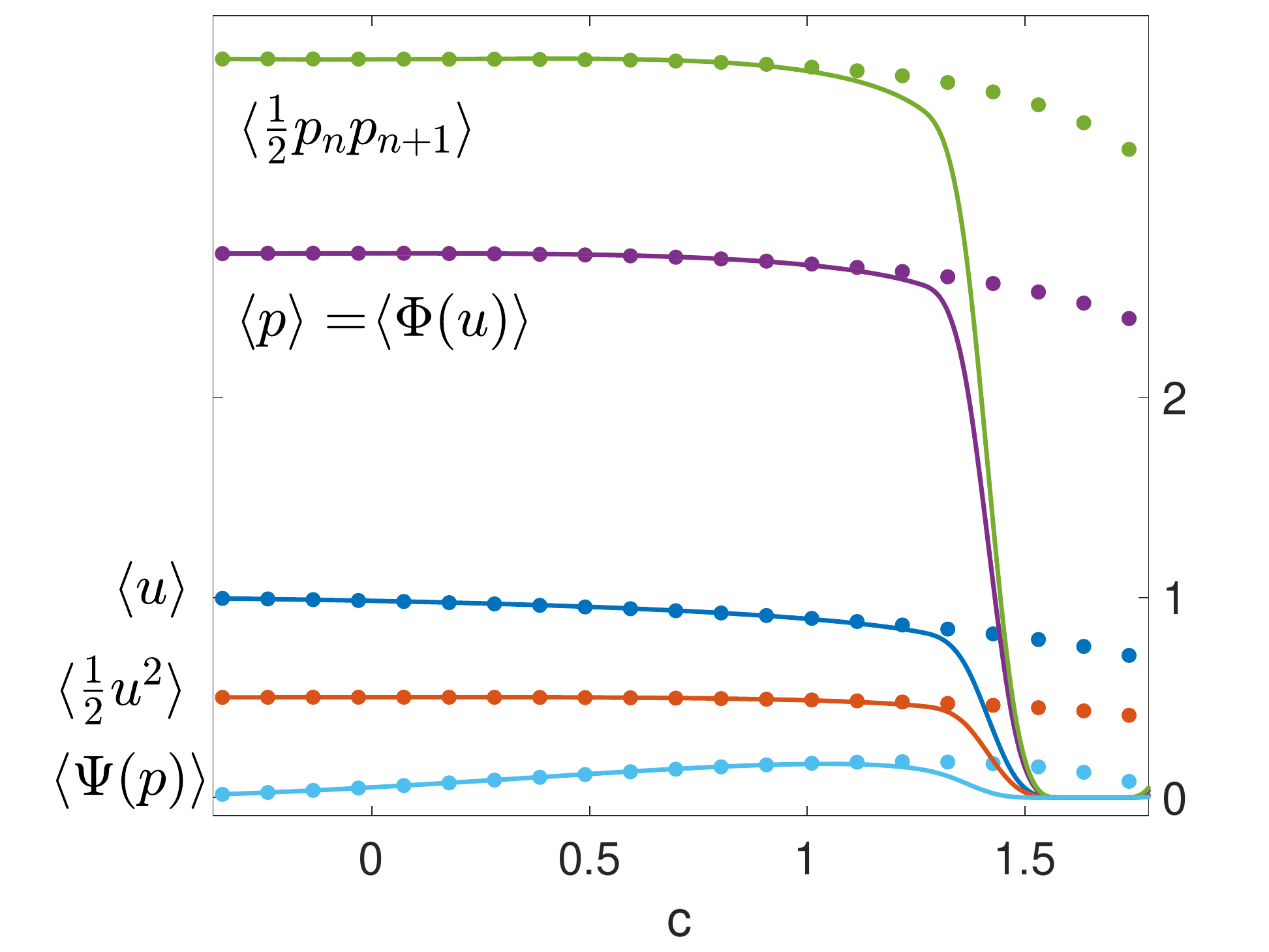}
  \end{tabular}  
 \caption{\textbf{(a)}
 Phase plane of the data extracted along the lines $c=X/T$ of a lattice DSW for the polynomial potential (markers), see also Fig.~\ref{fig:cTfix_data}(b).  
The solid lines are the planar ODE prediction of Eq.~\eqref{lowdim} using the energy
relationship given by Eq.~\eqref{kdvd_energy} (i.e., the quasi-continuum approximation).
 \textbf{(b)} Local averages measured from the lattice DSW (markers) and
 the ODE prediction based on the quasi-continuum approximation.
  \textbf{(c,d)} Same as panel (a) and (b), respectively, for the KvM potential.
  }
 \label{fig:qausi}
\end{figure*} 

There are various quantities that can be used to identify the character of a DSW. One is the spatial profile,
as shown in Fig.~\ref{fig:dsws}. For larger lattice sizes, however, the structure can be difficult to analyze
when viewing the DSW in this way. 
Another possibility is to inspect local averages, which can be computed directly from the lattice DSW data 
and from our ODE prediction. The local averages we consider are $<u_n> , <p_n>, <\Phi(u_n)>,
<p_n p_{n+1}/2>,<u_n^2/2>$ and  $<\Psi(p)>$ where the angle brackets denote the local average with respect to the traveling wave coordinate,
the dual variable is $p = \Phi'(u)$ and $\Psi(p) =p  u-\Phi(u)$ is the Legendre transform of $\Phi(u)$.
The reason we investigate these specific choices of local averages is their connection
to the modulation equations (see discussion in Sec.~\ref{sec:modulation}). 
Thus, we can bridge
results based on the standard modulation theory approach and our proposed ODE reduction approach.

Note that the local averages are, in principle, expressible in terms of the traveling wave parameters $V$, $K$ and $  \Omega$. For example,
$$\langle u\rangle :=\frac{1}{2 \pi}\int_0^{2 \pi}  V+\mathcal{V}(\zeta) \, d\zeta  = V, $$
where $\zeta$ is the traveling wave coordinate. The other local averages cannot be evaluated explicitly in general, however, since the periodic wave profile $\mathcal{V}$ is defined implicitly by an advance-delay equation, in which there is no closed-form solution. We estimate the local
averages numerically from the DSW data as a function of $c=n/t$ as follows
 \begin{equation} \label{eq:numerical_average}
     \langle u \rangle   \approx \frac{1}{T_p}\int_{n/c}^{n/c +T_p} \,  \frac{1}{2\mathcal{N}}\sum_{m=n-\mathcal{N}}^{n+\mathcal{N}}   u_m(t) \, dt. 
 \end{equation} 
where the sum appearing in the integrand is a mesoscopic average where $\mathcal{N}$ is chosen to be larger than
the spatial wavelength but much smaller than the lattice size. We chose $\mathcal{N} = 300$. 
The use of the mesoscopic spatial average leads to smoother data (e.g. the oscillations are effectively averaged out), but the data close to the boundaries will not have estimates for the averages.
Similar computations are made for the remaining five local averages (see the markers of
Fig.~\ref{fig:local}).

We will now compute local averages directly from the planar ODE solution of Eq.~\eqref{odesol}. For example,
\begin{equation} 
\langle u \rangle \, = \, \lambda_1  + \frac{\lambda_2-\lambda_1}{4K(m)} \Pi\left( \frac{\lambda_3-\lambda_1}{\lambda_3-\lambda_2},m\right), \label{mean}
\end{equation}
where $\Pi(C;m)$ is the complete integral of the third kind with elliptic characteristic $C$, and parameter $m$. 
The remaining five local averages can also be expressed in terms of elliptic
integrals (see the solid lines of Fig.~\ref{fig:local}).

The analytical predictions for the local averages based on the planar ODE are very close to the local averages computed directly from the lattice DSW for the polynomial potential 
and the KvM potential (compare the lines and markers of Fig.~\ref{fig:local}).  The analytical local averages provide a  layer of insight that is not available using the numerically obtained
averages. We were able to obtain the formulas
leveraging our data-driven approach, and
without the need to derive or solve a set 
of cumbersome modulation equations.

  \begin{figure*}
 \begin{tabular}{@{}p{0.25\linewidth}@{}p{0.25\linewidth}@{}p{0.25\linewidth}@{}p{0.25\linewidth}@{}  }
  \rlap{\hspace*{5pt}\raisebox{\dimexpr\ht1-.1\baselineskip}{\bf (a)}}
 \includegraphics[height=3.5cm]{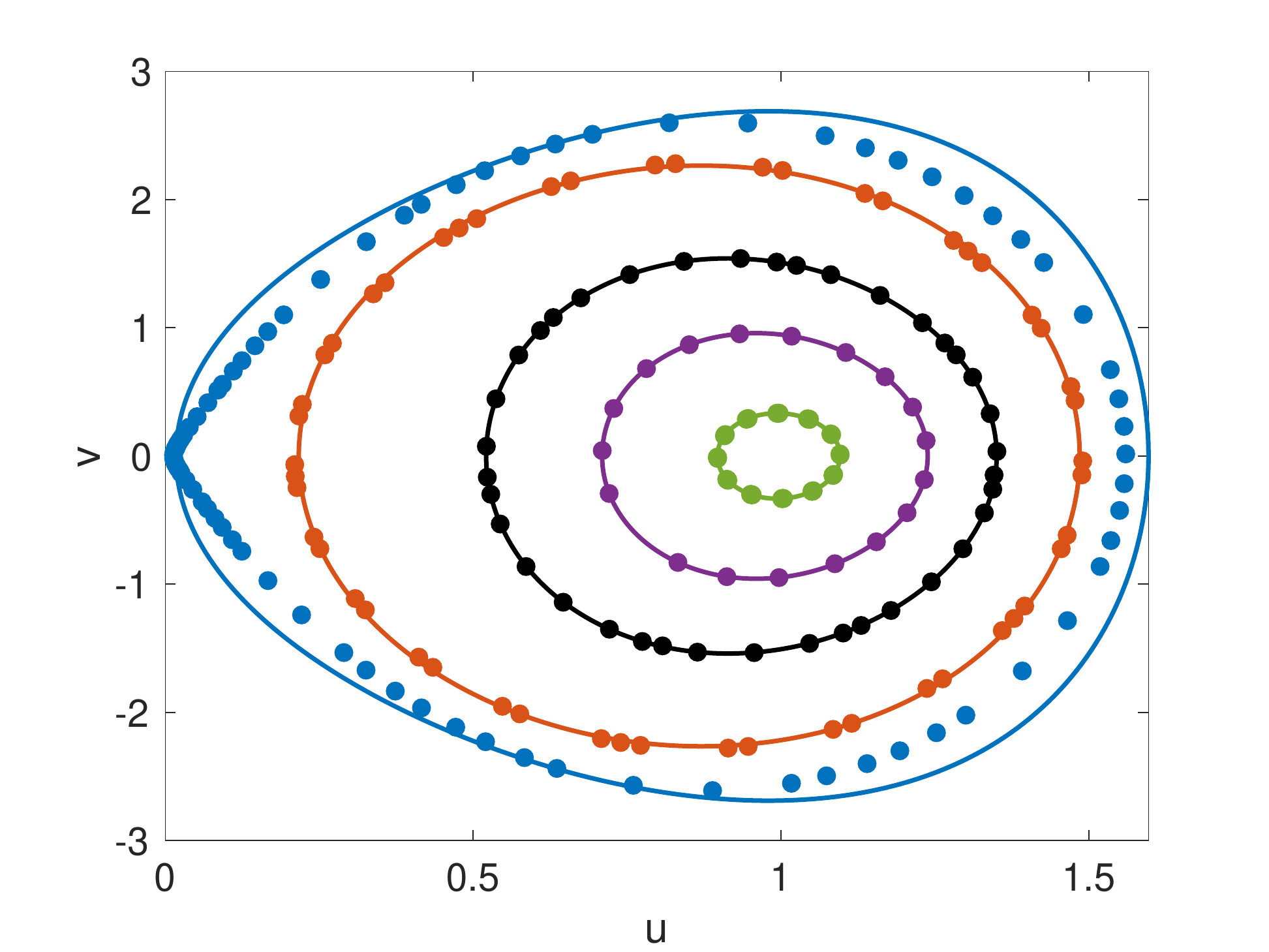} &
  \rlap{\hspace*{5pt}\raisebox{\dimexpr\ht1-.1\baselineskip}{\bf (b)}}
\includegraphics[height=3.5cm]{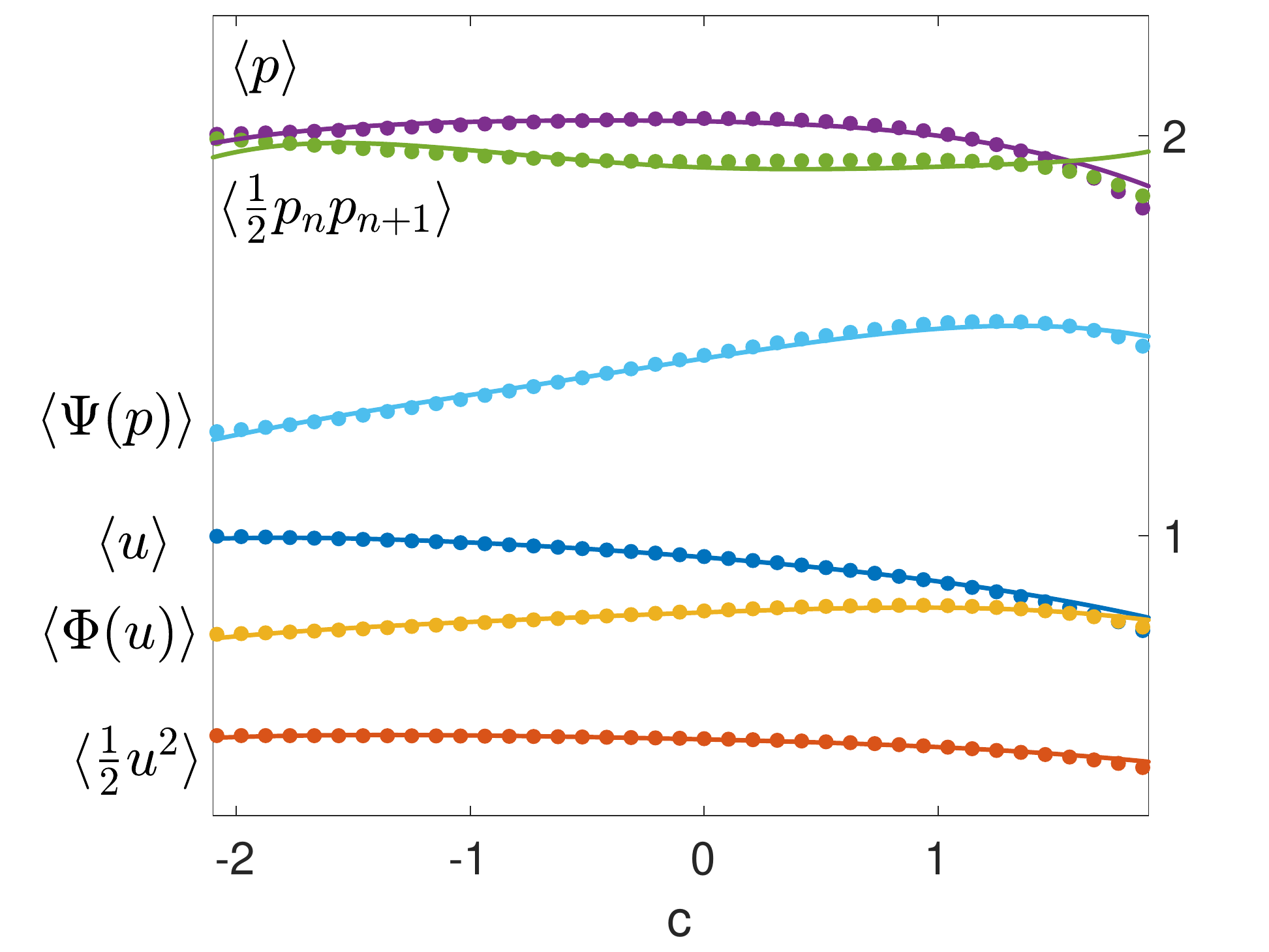}&
  \rlap{\hspace*{5pt}\raisebox{\dimexpr\ht1-.1\baselineskip}{\bf (c)}}
\includegraphics[height=3.5cm]{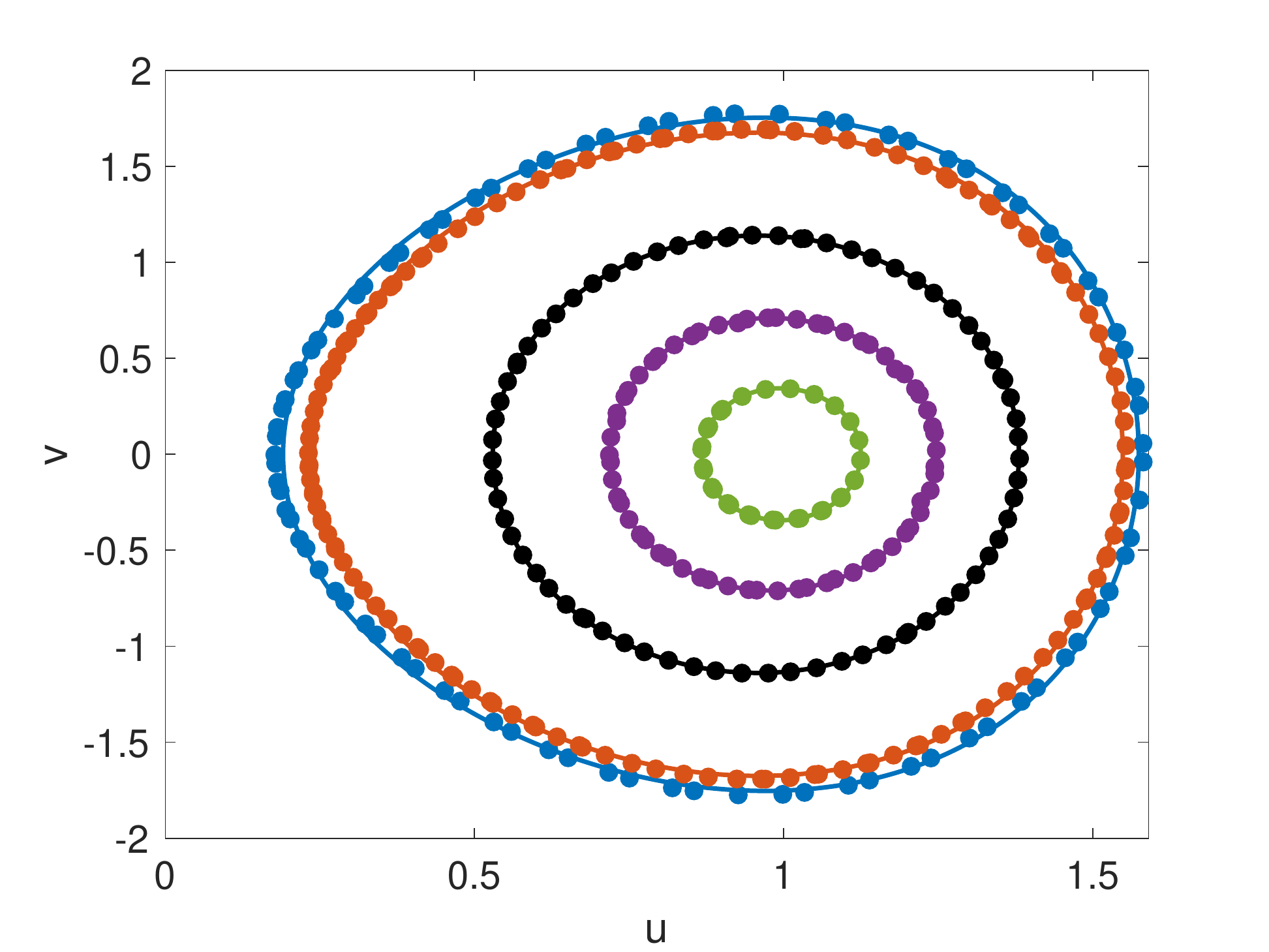} &
  \rlap{\hspace*{5pt}\raisebox{\dimexpr\ht1-.1\baselineskip}{\bf (d)}}
\includegraphics[height=3.5cm]{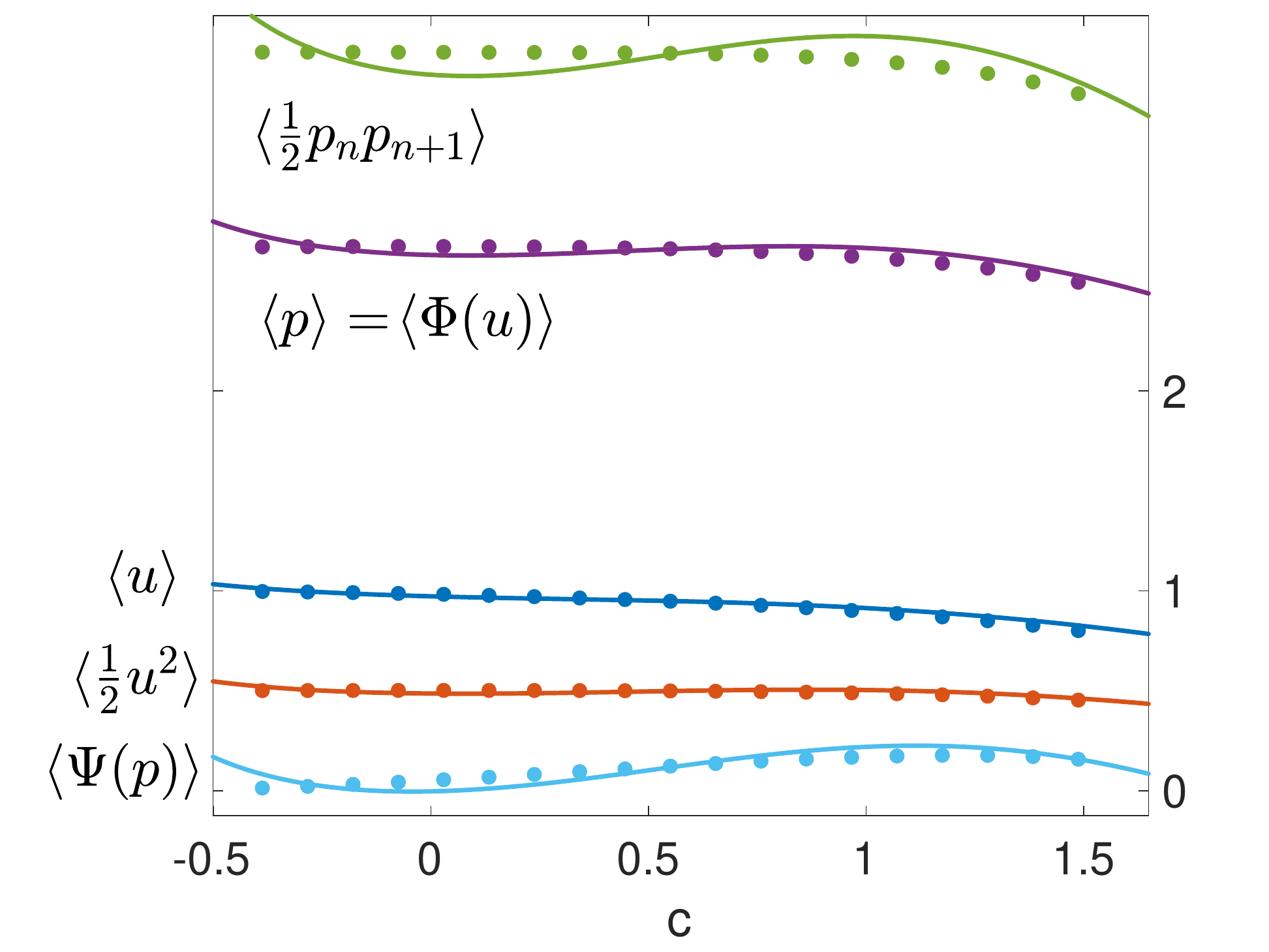}
  \end{tabular}  
 \caption{\textbf{(a)}
Same as Fig.~\ref{fig:qausi}, but using the regularized quasi-continuum reduction.
  }
 \label{fig:Rosenau}
\end{figure*}

\section{Quasi-continuum Approximation}\label{sec:quasi}

\subsection{A straightforward continuum limit}

 Having demonstrated that there is an underlying low-dimensional ODE structure to a lattice DSW,
it is reasonable to ask where this ODE structure comes from. While there are possibly many
ways in which this structure could manifest, a natural one to consider is the large lattice nature of the problem (in this article $N=8000$). We now investigate this possibility by taking a
suitable continuum limit of the lattice dynamics (bearing
a leading order lattice-induced correction). 
Using the hyperbolic scaling $u_n(t) = U(\epsilon n, \epsilon t) = U(X,T)$ the discrete conservation law
 Eq.~\eqref{deq} can be approximated by the following quasi-continuum model
\begin{equation}\label{KdVd}
\partial_T U = -\left(\partial_X+\frac{\epsilon^2}{6}\partial_X^3 \right) \Phi' (U),
\end{equation}
which is the same as Eq.~\eqref{eq} with the next order term in the expansion kept. Indeed, Eq.~\eqref{KdVd}
can be thought of as a dispersive regularization of Eq.~\eqref{eq}. Notice that the small parameter $\epsilon$ appears explicitly in the model equation,
making the error hard to control. This is why Eq.~\eqref{KdVd} is called a quasi-continuum
model rather than a continuum model. Uncontrollable errors are typical in quasi-continuum models, see the discussion in \cite{dcdsa},
which is in contrast to continuum models derived from other scalings, such as the Korteweg–de Vries or Nonlinear Schr\"odinger scalings, which have controllable errors \cite{SchneiderUeckerBook}.
Indeed, similar quasi-continuum models derived, for example, in the context of FPUT lattices have both failed and succeeded
in representing the actual lattice dynamics (see the discussion in \cite{pikovsky,granularBook}). Thus, it is important to check directly the validity of a PDE such as Eq.~\eqref{KdVd}. Here, we will do so at
the (reduced) level of the effective planar
ODE description. To that effect,
we first move into the co-traveling frame via the traveling wave ansatz $U(X,T) = \phi(K X -   \Omega T) = \phi(\zeta)$. 
Substitution of this expression into Eq.~\eqref{KdVd} and integrating once leads to
\begin{equation}
-  \Omega \phi  + \left (K+ \frac{\epsilon^2 K^3}{6}\frac{d^2}{d\zeta^2} \right ) \Phi' (\phi) + A= 0,
\end{equation}
where $A$ is an arbitrary integration constant. 
This equation has the following conserved quantity
$$E(\phi,v) = \frac{K^3\epsilon^2}{6} \frac{( \Phi''(\phi) v  )^2}{2}  + \int (K \Phi'(\phi) -   \Omega \phi + A) \Phi''(\phi) \, d\phi.   $$
If the planar structure identified in Fig.~\ref{fig:cTfix_data}(b) was a direct result of the large lattice size,
the trial form of the energy should be,
\begin{equation}
\frac{v^2}{2} =  \frac{6}{ \epsilon^{2} K^3} \frac{E - \int (K \Phi'(\phi) -   \Omega \phi + A) \Phi''(\phi) \, d\phi}{\Phi''(\phi)^2}.   \label{kdvd_energy} \end{equation}
 Using the functional form of Eq.~\eqref{kdvd_energy}, we find best-fit values of the parameters $E,K,  \Omega,A$
 as functions of $c$. The reduced ODE once again has the form given by Eq.~\eqref{lowdim}, but with $P$
 being replaced by the right-hand-side of Eq.~\eqref{kdvd_energy}. Solutions of this reduced ODE (computed numerically in this case)
 compare somewhat reasonably to the full lattice dynamics, see  Fig.~\ref{fig:qausi}(a,c). In this case, the deviation
 between the actual lattice DSW and the ODE solution becomes larger as the leading edge of the DSW is approached (see the outermost orbit of Fig.~\ref{fig:qausi}(a), for example). The difference between the ODE prediction and lattice DSW becomes even more apparent
 when inspecting the local averages, where the deviation can become significant for models with the polynomial potential and KvM potential,
see Fig.~\ref{fig:qausi}(b,d). In the case of the KvM potential, one notable difference when using Eq.~\eqref{kdvd_energy} instead of Eq.~\eqref{energy1}
is that the leading edge is predicted to occur for a smaller value of $c$. In particular the predicted leading edge is $\tilde{c}_+\approx 1.5$
where the observed leading edge is $c_+\approx 1.98$. This explains the steep decay of the local 
averages for $c\approx 1.5$ in Fig.~\ref{fig:qausi}(d). In both cases, it is clear
that the ODE prediction based on the quasi-continuum energy relationship yields poorer
results than those using the simple, data-driven polynomial energy relationship.
Nevertheless, the relevant prediction is quite adequate
for sufficiently small values of $c$ and provides a good
qualitatively and even reasonable semi-quantatively representation
of the corresponding phase portrait in Fig.~\ref{fig:qausi}
(see panels (a) and (c)), with a theoretical
foundation stemming from the underlying lattice
dynamics.
Before moving on, we briefly
investigate an alternative quasi-continuum model.

\subsection{The Rosenau regularization}

Quasi-continuum models derived in the fashion detailed above often have the unfortunate
side affect of being ill-posed due to large wavenumber instabilities in the associated dispersion relation \cite{Collins,Hochstrasse,Wattis93}.  In \cite{rosenau2,rosenau1} (see also the discussion in \cite{Nester2001}) Rosenau
proposed a regularization of such models to avoid the issue of ill-posedness. We will use the same regularization
in order to obtain an alternate quasi-continuum model. The regularized model is obtained
by inverting the operator $(1 + \frac{\epsilon^2}{6} \partial_X^2)$ in Eq.~\eqref{KdVd} 
and Taylor expanding it (on the left-hand side of the equation),  which leads to
$$ \partial_T U = \partial_X \Phi'(U) + \frac{\epsilon^2}{6} \partial_X^2 \partial_T U.  $$
Going into the co-traveling frame $U(X,T) = \phi(K X -   \Omega T) = \phi(\zeta)$ and integrating once, this PDE becomes the 
second order ODE
$$\frac{d^2 \phi}{d \zeta^2} = B + C \phi + D\Phi'(\phi),$$
where $B$ is an arbitrary integration constant and $C = 6/(   K^2 \epsilon^2)$ and $D = 6/(  \Omega K \epsilon^2)$.
This suggests the trial function for the energy should be
\begin{equation} \label{eq:Rosenau_energy}
    \frac{v^2}{2} = A + B u  + C \frac{u^2}{2} + D \Phi(u),
\end{equation} 
which has the advantage of being much simpler than Eq.~\eqref{kdvd_energy}.
In the case that the potential is a polynomial, namely $\Phi(u) = \frac{u^2}{2}+\frac{u^4}{4}$,
this trial energy is almost the same as the trial energy considered in Sec.~\ref{sec:datafit},
see Eq.~\eqref{energy1}. 
The only difference is the missing cubic term in 
Eq.~\eqref{eq:Rosenau_energy}. Thus, this trial function is more restrictive, but is "derivable",
and gives a physical meaning of the parameters of the fitting function in terms of
wave parameters $K,\Omega$. This is in contrast to the fit function used in Sec.~\ref{sec:datafit},
which was chosen out of the sake of simplicity.
Despite the slightly more restrictive nature of the trial energy, the results are reasonable
when comparing the ODE prediction to the full lattice DSW dynamics,
see Fig.~\ref{fig:Rosenau}. In both the polynomial and KvM potential, the Rosenau-type prediction is better than the non-regularized prediction (compare Figs.~\ref{fig:Rosenau} and \ref{fig:qausi}).  However, the Rosenau-type prediction is worse
than the prediction based on the general 4th order polynomial energy (compare Figs.~\ref{fig:Rosenau} and \ref{fig:local}).

The results of the non-regularized and regularized quasi-continuum predictions
suggest that the inherent underlying planar ODE structure is not due solely to the large lattice size of the problem, since the prediction based on
the more generic 4th order polynomial energy leads to the best results.
While further investigation of the  origin and 
structure of the planar ODE structure of the
lattice DSW lies beyond
the scope of the present article, it appears to
us to certainly be an issue
worthy of additional attention.

\begin{figure*}
     \begin{tabular}{@{}p{0.33\linewidth}@{}p{0.33\linewidth}@{} }
  \rlap{\hspace*{5pt}\raisebox{\dimexpr\ht1-.1\baselineskip}{\bf (a)}}
\includegraphics[height=4.5cm]{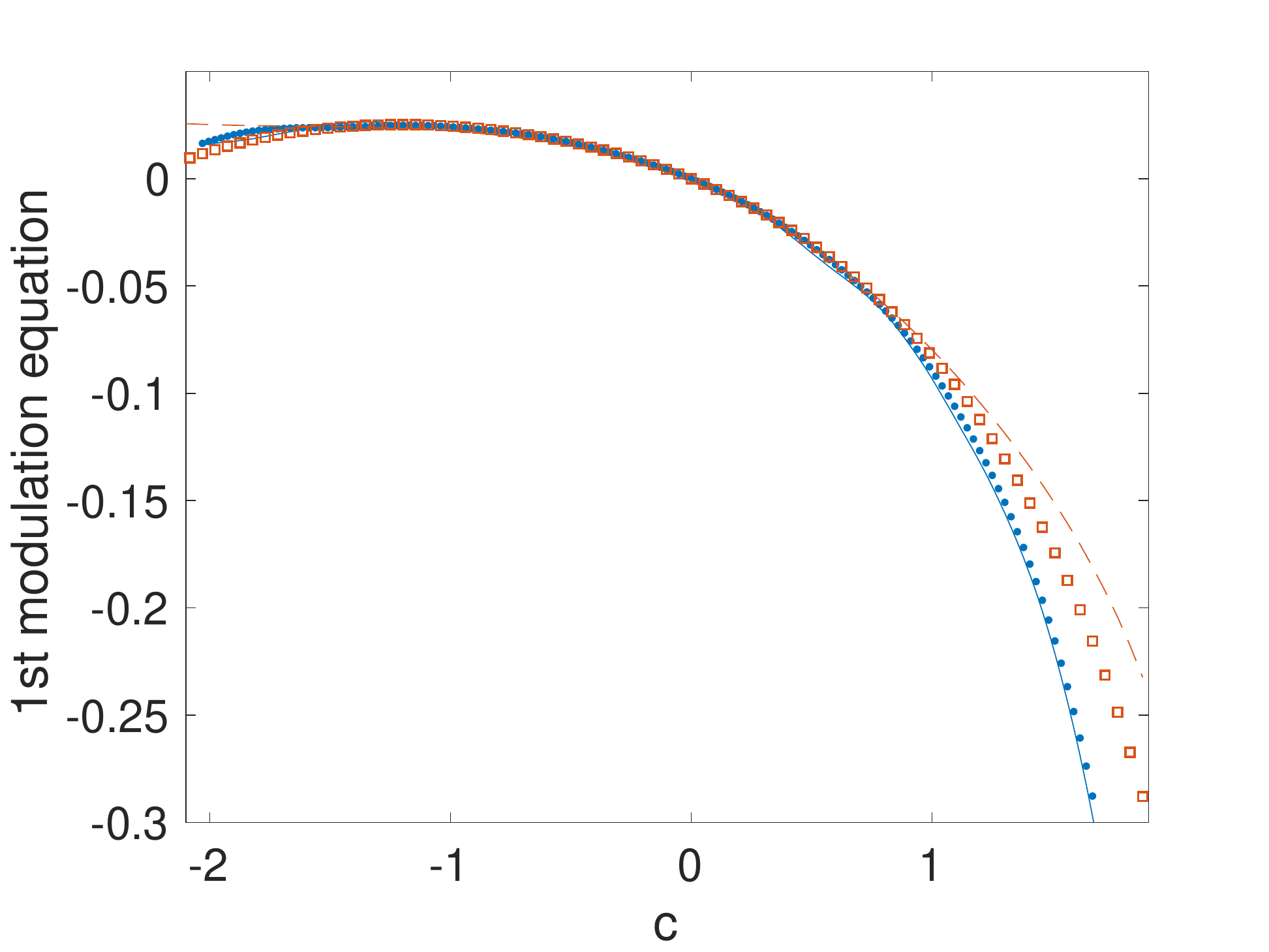}&
  \rlap{\hspace*{5pt}\raisebox{\dimexpr\ht1-.1\baselineskip}{\bf (b)}}
\includegraphics[height=4.5cm]{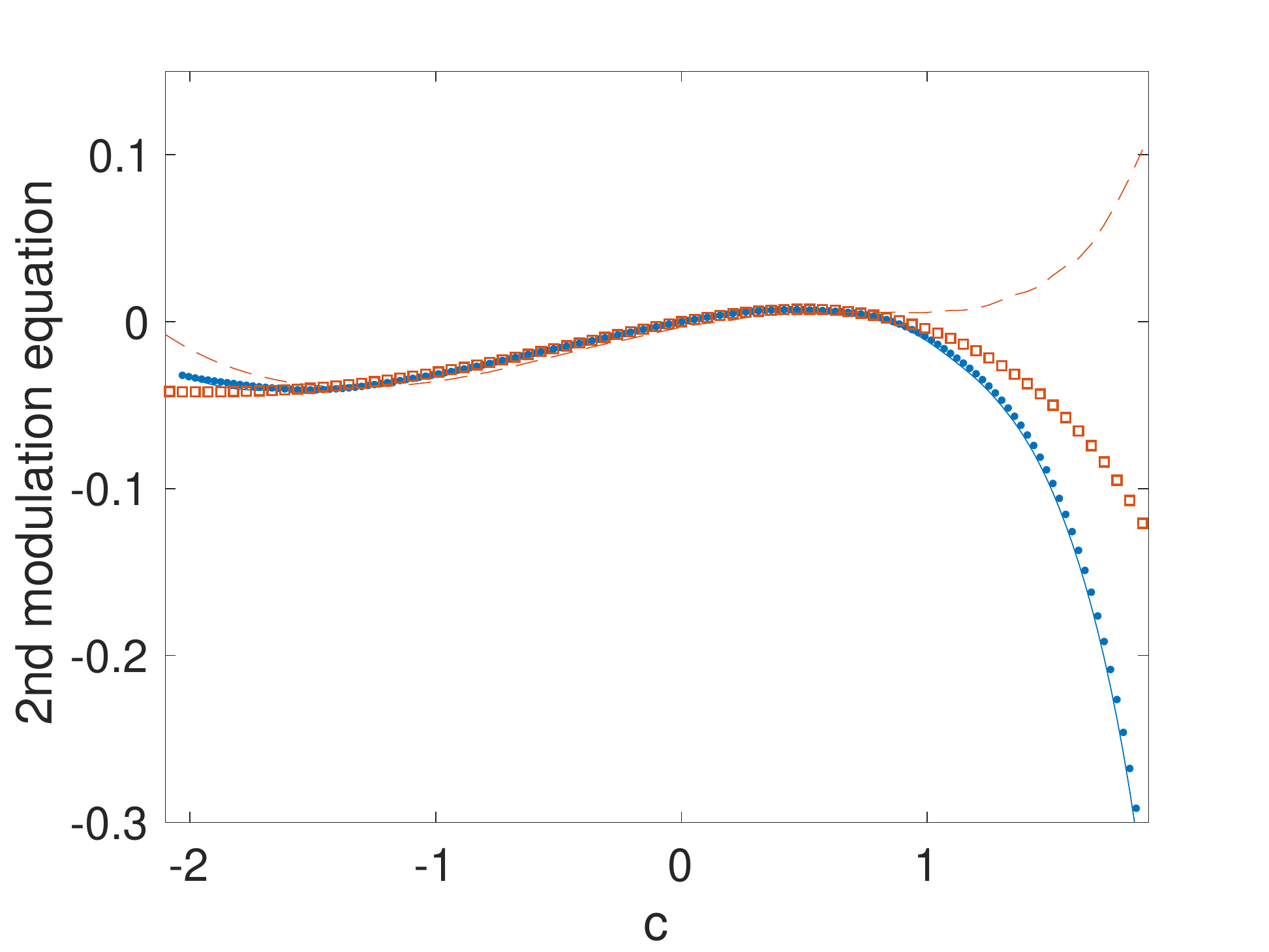}
  \end{tabular}
      \begin{tabular}{@{}p{0.33\linewidth}@{}p{0.33\linewidth}@{} }
  \rlap{\hspace*{5pt}\raisebox{\dimexpr\ht1-.1\baselineskip}{\bf (c)}}
\includegraphics[height=4.5cm]{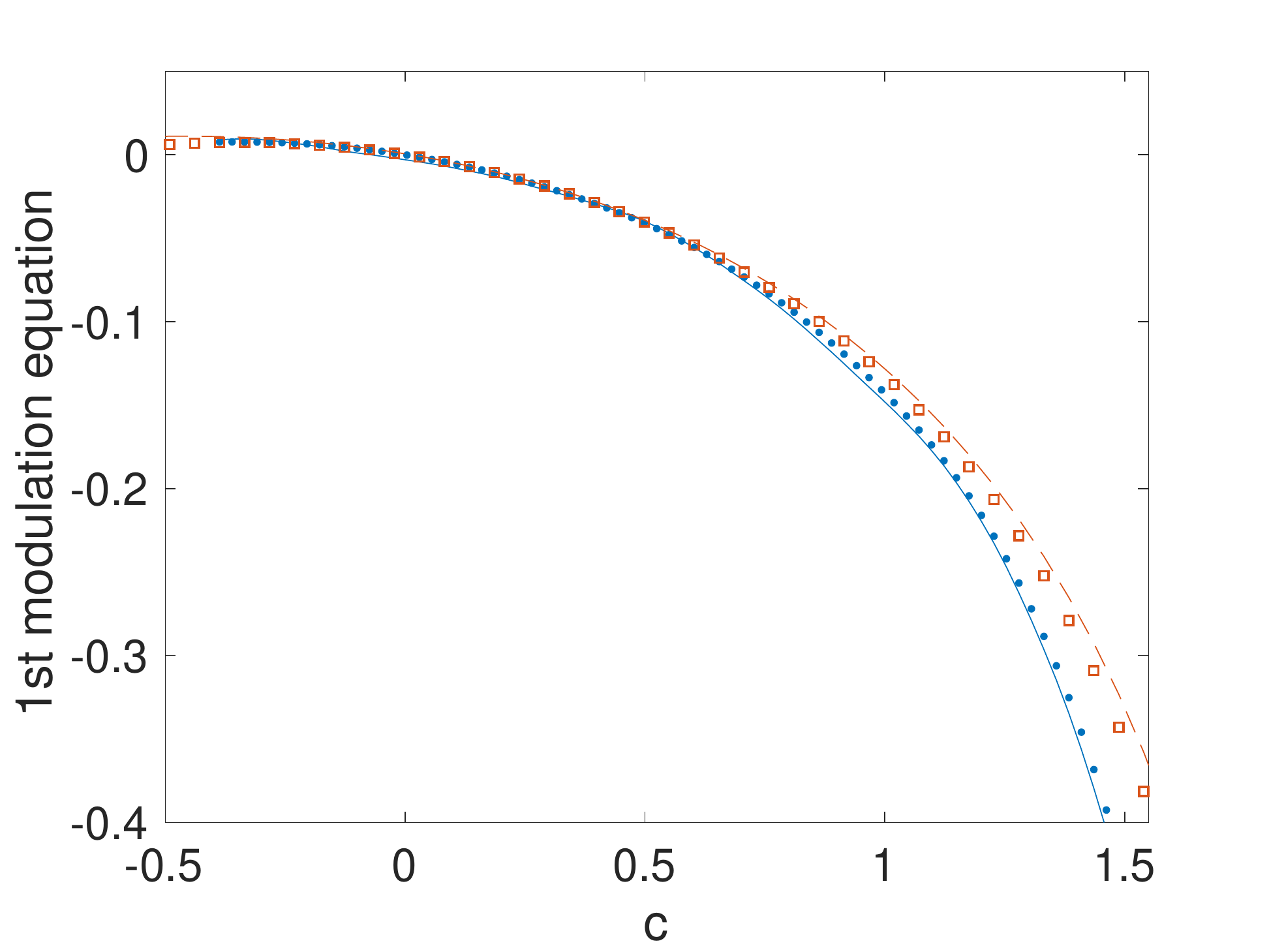}&
  \rlap{\hspace*{5pt}\raisebox{\dimexpr\ht1-.1\baselineskip}{\bf (d)}}
\includegraphics[height=4.5cm]{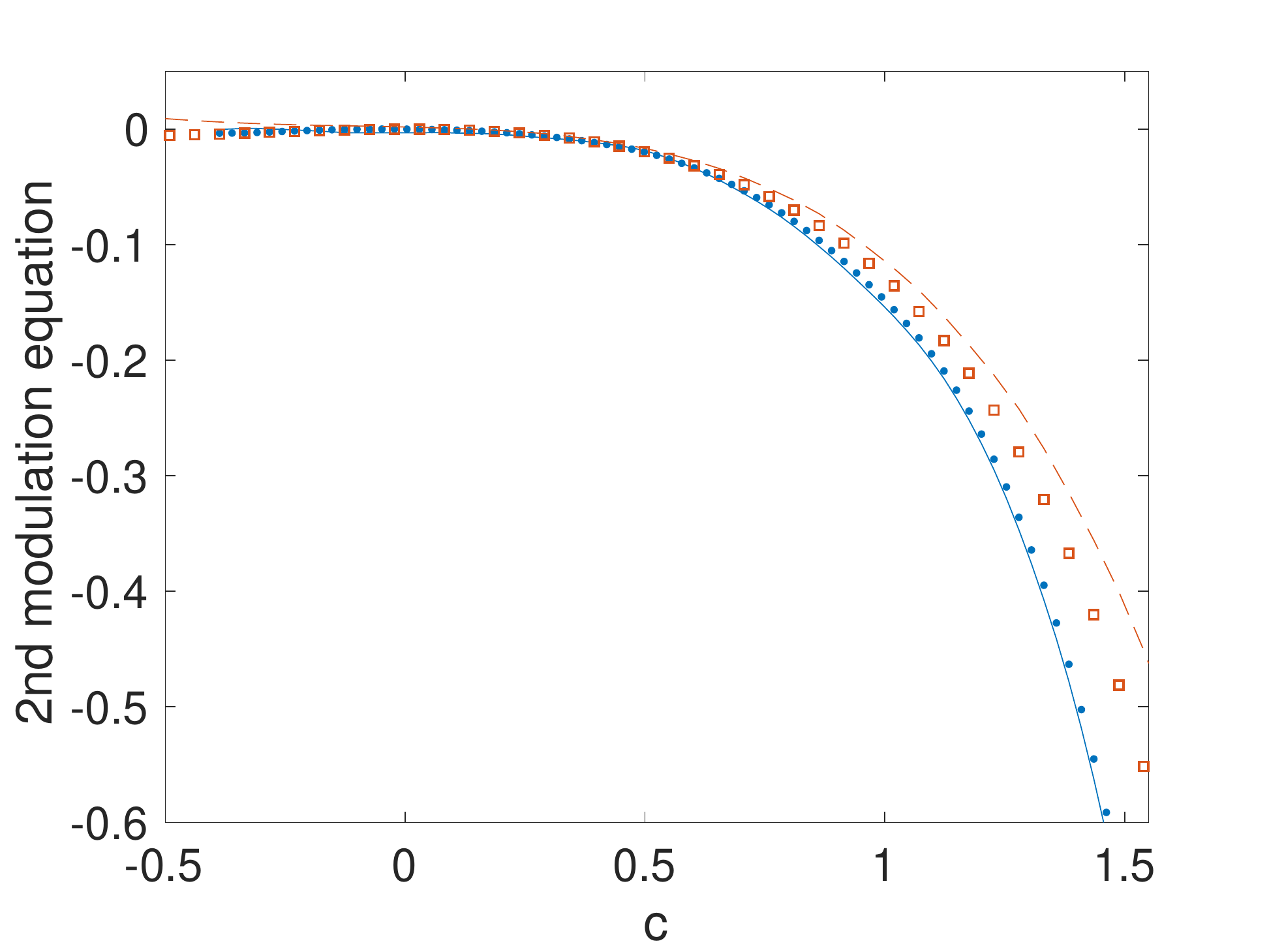}
  \end{tabular}
 \caption{\textbf{Top row:} Comparison of ODE predictions and full lattice DSW dynamics
 for the polynomial potential $\Phi(u) = u^2/2 + u^4/4$.
 \textbf{(a)} Plot of the quantity $\frac{\mathrm{d}}{\mathrm{d} c}\langle  p_n
\rangle$
 for the lattice DSW (blue solid line) and the ODE prediction (red dashed line). Plot of
 the quantity $c \frac{\mathrm{d}}{\mathrm{d} c}\langle u_n
\rangle$
for the lattice DSW (blue points) and the ODE prediction (red squares).
 \textbf{(b)} Plot of the quantity $\frac{\mathrm{d}}{\mathrm{d} c}\langle p_n p_{n+1} / 2  \rangle$
 for the lattice DSW (blue solid line) and the ODE prediction (red dashed line).
 Plot of the quantity $c \frac{\mathrm{d}}{\mathrm{d} c}\langle \Phi(u_n) \rangle$
for the lattice DSW (blue points) and the ODE prediction (red squares).
\textbf{Bottom row}:  Same as the top row with the KvM potential $\Phi(u) = \exp(u)$.
}
 \label{fig:modulation}
\end{figure*}

\section{Modulation Theory} \label{sec:modulation}

We will now briefly describe the modulation equations from the perspective of local conservation laws, and discuss the comparison thereof with the results obtained
using the planar ODE reduction. For the sake of completeness, 
the derivation of the modulation equations in terms of the wave parameters using
Whitham's method of the averaged Lagrangian is included in Appendix \ref{app:mod}.

One can derive modulation
equations by averaging local conservation laws of the system \cite{Whitham74,Mark2016}.
Upon evaluation of the local averages, one can express the equations
in terms of the original wave parameters, such as mean amplitude, modulation wavenumber, etc.
If the equations are tractable, one solves the modulation equations
in a self-similar frame subject to boundary data that are consistent with the DSW
(namely that the states $u_-$ and $u_+$ are connected
via the modulated wave train). For a detailed discussion of this procedure see \cite{Gurevich1987,Kamchatnov}.

 The equation of motion \eqref{deq} implies the discrete energy law
\begin{equation*} 
\frac{d}{dt}\Phi(u_{n} ) + \left(\tfrac12\,\Phi'(u_{n} )\,\Phi'(u_{n+1} )\right)-\left(\tfrac12\,\Phi'(u_{n-1} )\,\Phi'(u_{n} )\right)= 0, 
\end{equation*}
and we observe that both equations are discrete counterparts of conservation laws since the time derivative of a microscopic observable equals a discrete spatial derivative of another quantity. Using the concept of weak convergence, we can pass to a continuum limit and obtain the weak formulation of the macroscopic PDE  
\begin{equation} \label{mde}
\begin{array}{lll}\partial_{T}\langle u_n\rangle & +  \partial_{X}\left\langle p_n  \right\rangle & =0, \\
 \partial_{T}\langle\Phi(u_n)\rangle & +  \partial_{X}\left\langle\frac{1}{2} p_n p_{n+1} \right\rangle & =0, 
 \end{array}
\end{equation}
where $p_n=\Phi'(u_n)$ and the angle-brackets denote macrosopic fields that depend on $X$ and $T$. More precisely, denoting by $\varphi$ a smooth and compactly supported testfunction we have
\begin{align*}
\sum_{n}\!\int \!\!\frac{d}{d t}u_n(t)\,\varphi(\eps n,\eps t)\to-\iint
\langle u_n\rangle(X,T)\,\partial_T\varphi(X,T) d\!Xd\!T,
\end{align*}
as $\eps\to0$ thanks to the scaling relations $T=\eps t$, $X=\eps n$
and integration by parts with respect to time. The field $\langle u_n\rangle$ represents the weak limit
of $u_n$ and its value at any point $(X,T)$  can be viewed as the mean value of $u_n(t)$ in a mesoscopic space-time window centered around that point. Similar arguments apply to discrete spatial derivatives and to other observables, so \eqref{mde} is an immediate consequence of the lattice dynamics. For a space-time continuous conservation law system,
translation invariance yields another conservation law (momentum) via Noether's theorem. Following
this naively for the lattice system, one finds the equation
$$\partial_{T}\left\langle\frac{1}{2} u^{2}\right\rangle  +\partial_{X}\left\langle \Psi(p) \right\rangle  =0,$$
where $\Psi(p) =p  u-\Phi(u)$ is the Legendre transform of $\Phi(u)$. 
In the lattice setting, however, there is no translation invariance, and hence, no corresponding conserved quantity.
For this reason, one cannot expect this third modulation equation to be valid in the lattice setting and, thus,
the set of modulation equations is incomplete. A complete set of three modulation equations is 
discussed in Appendix \ref{app:mod}. Since our goal is not to use modulation equations to obtain a description
of the DSWs (although we recognize that the latter
is the standard approach towards their description
to which we offer an alternative herein), the missing modulation equation is not critical to our discussion.
Indeed, the proposed ODE reduction approach allows us to avoid the modulation equations entirely, 
and they are presented here only to demonstrate that our ODE reduction is consistent with the modulation theory.

We claimed the modulation Eqs.~\eqref{mde} are valid in the lattice setting, a feature that
we now further discuss. A DSW will have a self-similar structure and thu, it is useful to define the self-similar coordinate $c = X/T$
in which case the modulation equations become
\begin{eqnarray} \label{mde2}
  \frac{d}{dc}\left\langle p_n  \right\rangle & =& c \frac{d}{dc}\langle u_n \rangle ,  \\
   \frac{d}{dc}\left\langle\frac{1}{2} p_n p_{n+1}\right\rangle & =& c \frac{d}{dc}\langle\Phi(u_n)\rangle .
\end{eqnarray}
The first modulation equation is inspected by plotting the quantities $\frac{\mathrm{d}}{\mathrm{d} c}\langle  p_n \rangle$ and 
$c \frac{\mathrm{d}}{\mathrm{d} c}\langle  u_n\rangle$ as functions of $c$,
see the blue lines and markers of Fig.~\ref{fig:modulation}(a), where overlapping curves correspond to the modulation equation being valid.
The validity of the second modulation equation is demonstrated in Fig.~\ref{fig:modulation}(b), where
the quantities $\frac{\mathrm{d}}{\mathrm{d} c}\langle p_n p_{n+1} / 2  \rangle$
and $c \frac{\mathrm{d}}{\mathrm{d} c}\langle \Phi( u_n ) \rangle$ are shown as functions of $c$,
see the blue lines and markers of Fig.~\ref{fig:modulation}(b). The ODE predictions of these quantities 
are consistent with the modulation equations for regions sufficiently far from the leading edge
of the DSW, see the red dashed lines and squares of Fig.~\ref{fig:modulation}(a,b). There is noticeable
deviation of the ODE prediction corresponding to the second modulation equation 
for $c\approx 1.1$, and it is found to be larger in the case of the polynomial potential. In particular, the ODE prediction (red dashed line) becomes positive while in the full lattice dynamics the relevant quantity is negative. This large deviation
stems from small deviations of $\langle p_n p_{n+1} / 2  \rangle$ due to
changes in slope. In particular, in Fig.~\ref{fig:local}(a) 
note how the ODE
prediction (green line) is increasing, whereas, the 
actual lattice DSW quantity (green markers) is decreasing.  This observation is consistent with general deviations of the ODE orbits
and full lattice DSW orbits at the edges of the DSW discussed previously, with
the deviations in terms of derivatives of the local averages being more noticeable. 
 Overall, however,  Fig.~\ref{fig:modulation} demonstrates
that the ODE reduction is consistent with the modulation equations
for regions of the DSW sufficiently far from the leading and trailing edges.
The results for the KvM potential are similar and even slightly better, see Fig.~\ref{fig:modulation}(c,d). Note that the second modulation
equation is also consistent with the ODE prediction, even for wave speeds
close to the leading edge, as shown in Fig.~\ref{fig:modulation}(d).

\section{Conclusions \& Future Challenges} \label{sec:theend}

In the present work, we have demonstrated that there is an underlying low-dimensional structure within the core of a lattice DSW.
The standard approach based on Whitham modulation theory is  
for most purposes impractical in the context of lattice DSWs. 
For this
reason, we have argued about the relevance of elucidating
and identifying either via a data-driven or in an 
approximate (quasi-continuum) analytical
way the form of this effective planar ODE description of the
lattice DSWs. We have argued that  this
approach can
yield the kind of information typically inferred from  the modulation equations, but it is much simpler in its
implementation, given that it only requires the analysis
of a planar ODE. 
Indeed, in principle, our data-driven approach, which
has proved most efficient in the numerical examples
considered herein, only
needs a suitable manipulation of the data set 
stemming from a (e.g., potentially black-box) integrator
in order to be set up.
We believe that the
present findings may help lay the groundwork for further studies that adopt such a low-dimensional
approach towards the study of lattice DSWs as an alternative to modulation theory, including in a variety of different models (such as, e.g., in discrete
models
of the nonlinear Schr{\"o}dinger variety~\cite{Kamchatnov2004,PK09}). For example,
we showed that the most natural candidate for the underlying 
ODE based on the quasi-continuum approach
did not perform as well as the simple polynomial approximation.
We also considered an alternative variant via the
Rosenau regularization which was deemed to be more effective
than the former but less so than the latter (yet was also systematically
possible to obtain from the microscopic model). The
different approaches 
suggest there may be other derivable ODEs
waiting to be discovered and exploited in a manner similar to that presented here. On a broader (and more challenging)
scale, a potential self-consistent, general formulation of the Whitham modulation
equations for nonlinear dynamical lattices 
remains an important, but highly challenging task as was highlighted
herein. Arguably, this is, to a nontrivial degree,
due to the
lack of some symmetries (most notably of translation
invariance) in the lattice setting. Studies along 
these directions are currently underway and will
be reported in future publications.

\section*{Acknowledgement}
The present paper is based on work that was supported by the US National Science Foundation under Grant No. DMS-2107945 (CC)
and DMS-1809074 (PGK).

\appendix
\section{Linear Tail Size} \label{app:tail}

\begin{figure*}
 \begin{tabular}{@{}p{0.25\linewidth}@{}p{0.25\linewidth}@{}p{0.25\linewidth}@{}p{0.25\linewidth}@{}  }
  \rlap{\hspace*{5pt}\raisebox{\dimexpr\ht1-.1\baselineskip}{\bf (a)}}
 \includegraphics[height=3.5cm]{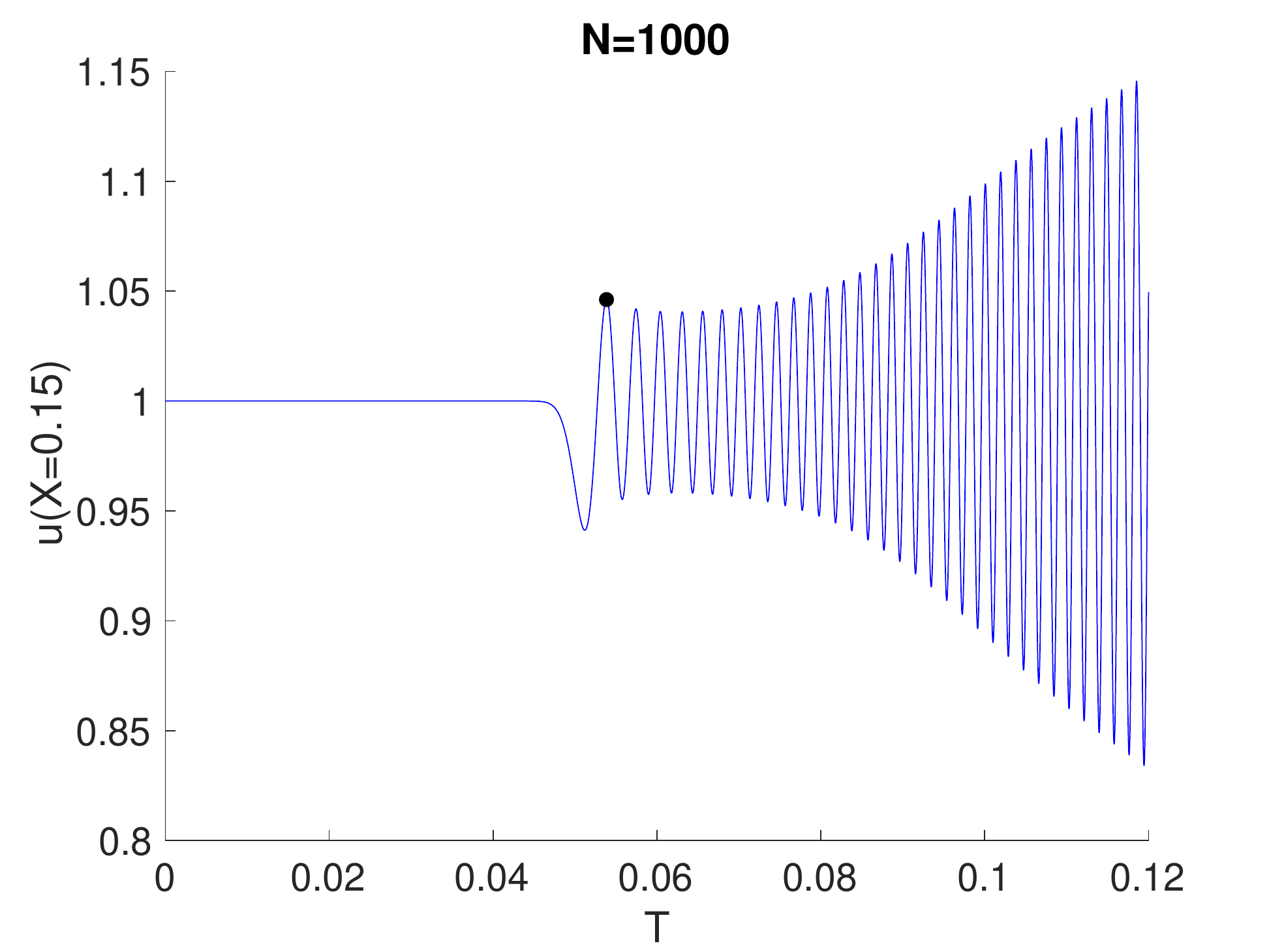} &
  \rlap{\hspace*{5pt}\raisebox{\dimexpr\ht1-.1\baselineskip}{\bf (b)}}
\includegraphics[height=3.5cm]{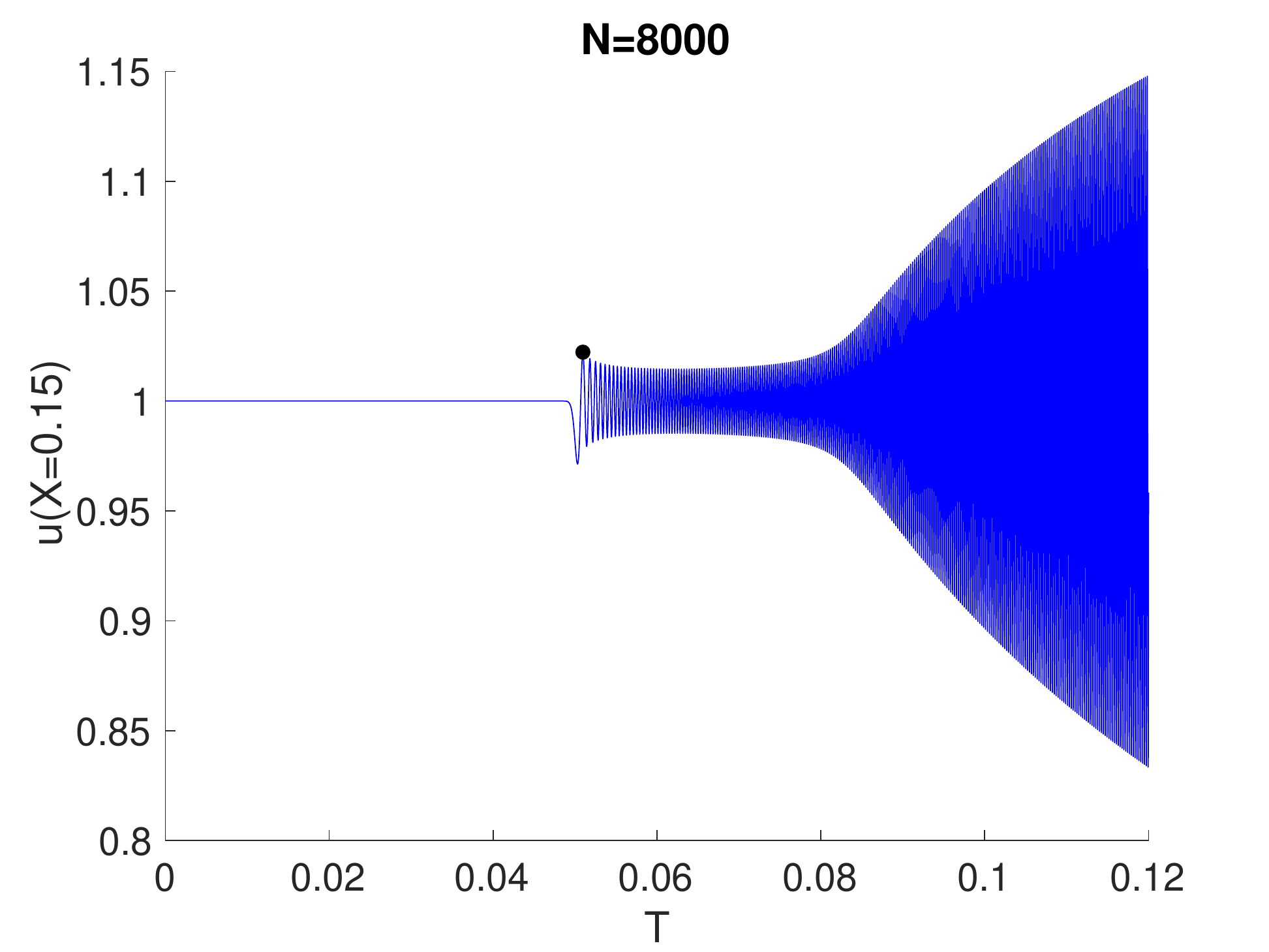}&
  \rlap{\hspace*{5pt}\raisebox{\dimexpr\ht1-.1\baselineskip}{\bf (c)}}
\includegraphics[height=3.5cm]{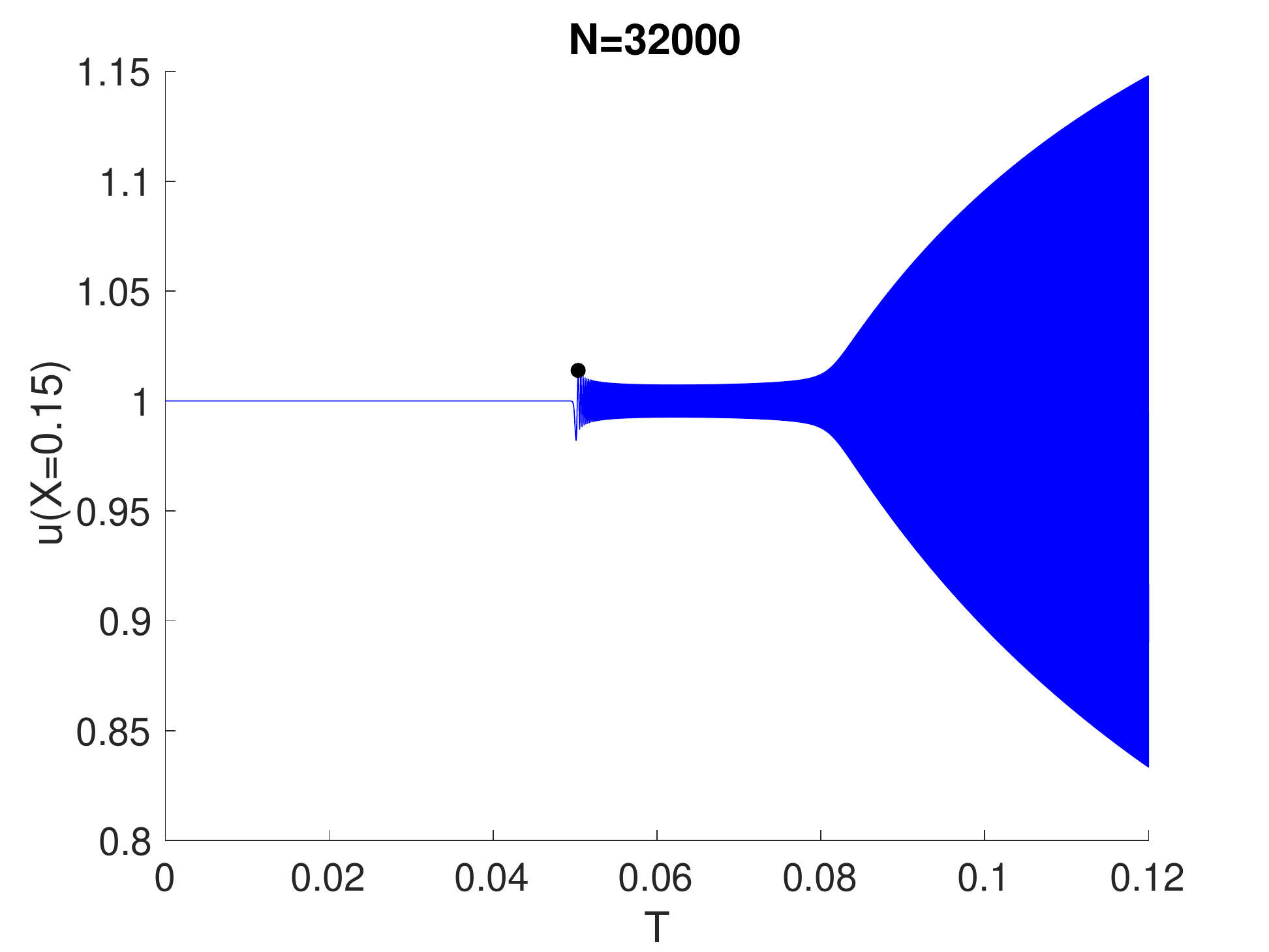} &
  \rlap{\hspace*{5pt}\raisebox{\dimexpr\ht1-.1\baselineskip}{\bf (d)}}
\includegraphics[height=3.5cm]{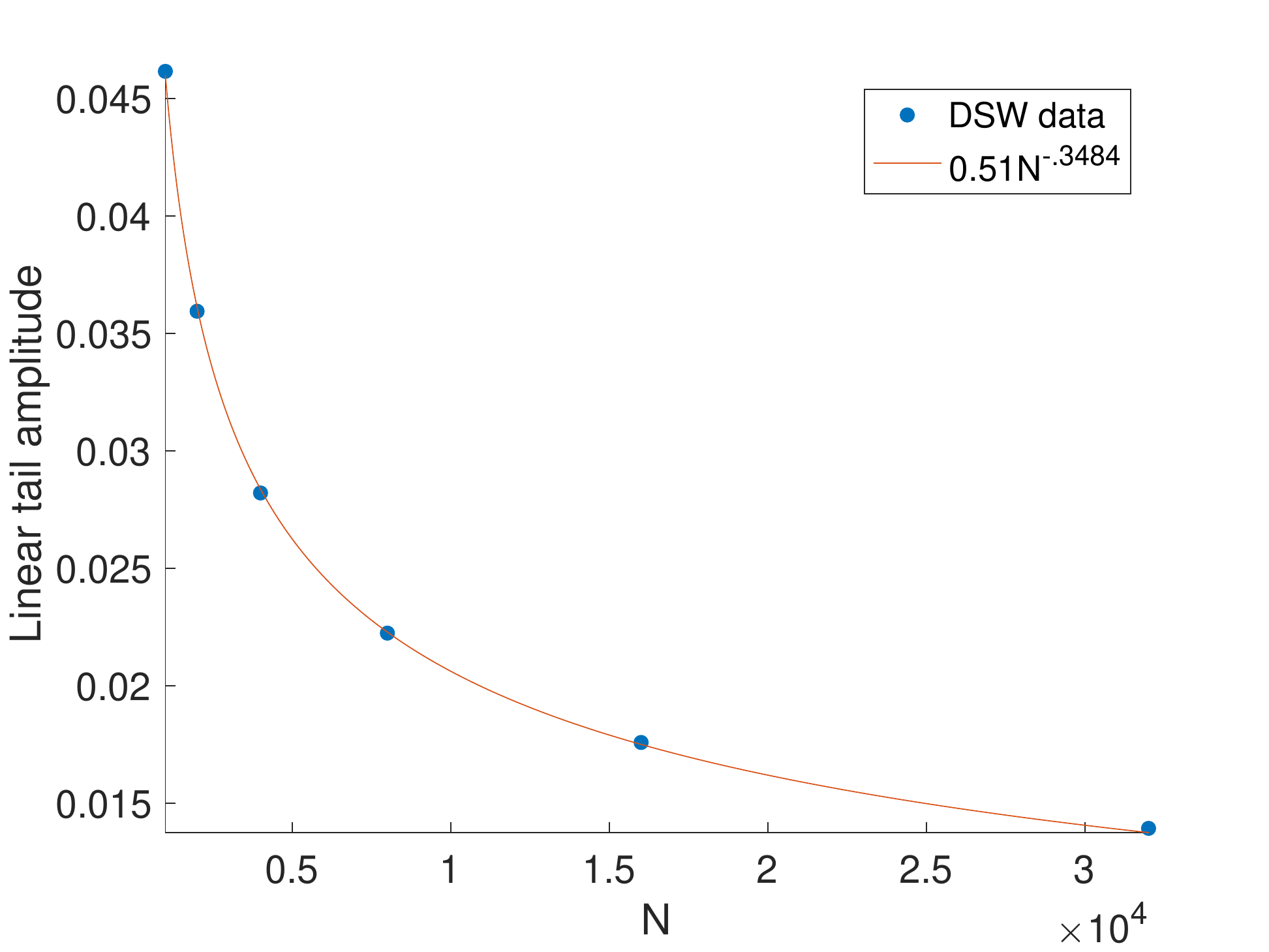}
  \end{tabular}  
 \caption{Time series data for fixed $X = 0.15$ for the lattice sizes \textbf{(a)} $N=1000$, \textbf{(b)} $N=8000$ and \textbf{(c)} $N=32000$. The black dot is taken
 as the reference point to measure the amplitude of the linear wave (first local maximum of the linear wave). \textbf{(d)}
 Plot of the amplitude of the linear wave (i.e. the black points) against the lattice size $N$. The red curve is the best-fit curve
 of the form $a N^{b}$, where $a=0.51$ and $b = -0.3484$, which is consistent with the linear theoretical prediction
 that the decay law is $\sim N^{-1/3}$. }
 \label{fig:tails}
\end{figure*}

 According to the linear theory, the amplitude of the linear wave at the trailing edge of the DSW can be shown to follow the decay law $\sim N^{-1/3}$
 with the use of Fourier analysis \cite{MielkePatz2017}. In Fig.~\ref{fig:tails}, time series plots are shown for three lattice sizes, $N=1000$, $N=8000$ and $N=32000$ where it is seen
that the amplitude of linear waves is decreasing. In particular, the time series for the $X = 0.15$ node is shown, where $X = \epsilon n$ and $\epsilon = 1/N$. In these graphs, the slow variable is defined as $T = \epsilon t$. The last panel shows
 a plot the amplitude of the linear wave  against the lattice size $N$ where it is seen that the observed decay is consistent with the theoretical decay law $\sim N^{-1/3}$. 

\section{Modulation Equations in Terms of Wave Parameters} \label{app:mod}

For the sake of completeness, we present the modulation equations in terms of the wave parameters
using Whitham's method of the averaged Lagragian. For the lattice, this procedure yields a complete set of three modulation equations, but, as the derivation will demonstrate, working with them is cumbersome. 

The formal procedure to obtain Whitham modulation equations in the lattice setting is well-known \cite{Venakides99,DHM06}. Although rigorous error estimates in lattice or continuous settings are rare (see \cite{SchnDull2009,DHM06} for special examples), numerical simulations do suggest they are valid \cite{DH08}. 
Before
writing the modulation equations, we must re-write the traveling wave equation  --- and actually  also the lattice --- 
using an integral formulation. We start by recalling that traveling periodic waves of Eq.~\eqref{deq} have the form  $u_{n}(t)=V+\mathcal{V}(\zeta), \, \zeta = K n-  \Omega t$, where the parameters $V$ (mean), $K$ (wavevector) and $  \Omega$ (frequency) are real and
 the wave profile $\mathcal{V}$ is assumed, w.l.o.g, to be $2\pi$ periodic with mean zero. 
The traveling wave equation has a variational structure, which will be exploited to derive modulation
equations. To see this, we define the centered difference operator $\nabla_K$ (which is skew-symmetric) by
\begin{align*}
 \left( \nabla_K \calF \right) (\zeta) = \calF( {\zeta+K} ) -\calF({\zeta -K }),
\end{align*}
and define $\calW$ via
\begin{align}
\label{Eqn:RelVW}
\calV = \nabla_K\calW\,.
\end{align}
Note  that the traveling wave equation \eqref{adv1} is just the Euler-Lagrange equation 
\begin{align*}
\partial_{\calW} \calL\quadruple{\calW}{V}{K}{ \Omega}=0,
\end{align*}
to the  Lagrangian functional
\begin{equation} \label{eq:lagragian}
    \calL\quadruple{\calW}{V}{K}{ \Omega} = \Omega \mathcal{S}(\mathcal{W}, K) - \mathcal{E}(\mathcal{W}, V, K),
\end{equation}
where 
\begin{equation}
\label{Eqn:SFunctional}
\mathcal{S}(\mathcal{W}, K)= \frac{1}{2\pi}\int_0^{2 \pi} \left(\nabla_{K} \mathcal{W}\right)(\zeta) \cdot \mathcal{W}^{\prime}(\zeta) \mathrm{d} \zeta,
\end{equation}
and
\begin{equation}
\label{Eqn:EFunctional}
\mathcal{E}(\mathcal{W}, V, K)=\frac{1}{2\pi}\int_0^{2 \pi}\Phi\left(V+\left(\nabla_{K} \mathcal{W}\right)(\zeta)\right) \mathrm{d} \zeta.
\end{equation}
The traveling wave equation \label{adv} can then be written as
\begin{align} \label{conopt}
  \Omega\,\partial_{\calW}\calS\pair{ \calW }{K}=\partial_{\calW}\calE\tripple{\calW}{V}{K}\,,
\end{align}
where $ \Omega$ plays the role of a Lagrange multiplier. For the linear model,
$\Omega$ is related to $K$ through the dispersion relationship,  but for nonlinear functions $\Phi^\prime$, we expect that $V$, $K$ and $\Omega$ are independent parameters. The existence
of a three-parameter family of traveling wave solutions to Eq.~\eqref{deq} was demonstrated in \cite{Herrmann_Scalar}, albeit with different parameters and using a certain reformulation of \eqref{adv1}.

 The DSW can be thought of
as a traveling wave, as just described, but now the wave parameters $V$, $K$ and $\Omega$ vary in time and space. 
This motivates an ansatz for an approximate modulated wave solution,
\begin{equation}\label{mtw}
 w_{n}(t) \approx \frac{1}{2\,\epsilon}\,A(X,T)+\mathcal{W}\left(X,T; \frac{1}{\epsilon}Z(X,T)\right),
\end{equation}
where the modulated traveling  waves parameters are given  by
\begin{align*}
V=\partial_X A\,,\qquad \Omega=-\partial_T Z 
\,,\qquad  K=\partial_X Z\,. 
\end{align*}
If one substitutes ansatz \eqref{mtw} into the total action integral for Eq.~\eqref{deq}, that is
\begin{align*}
\mbox{total action}&=-\int\sum_{n} \dot{w}_n(t)\big(w_{n+1}(t)-w_{n-1}(t)\big) dt
\\&\quad\quad\quad -
 \int\sum_{n}\Phi\big(w_{n+1}(t)-w_{n-1}(t)\big)  dt   \,,
\end{align*}
and replaces summation by integrals, one arrives at
the expression:
\begin{align*}
    \mbox{total action}&= -\tfrac{1}{2} \int\limits_0^{T_f}  \int\limits_{X_0}^{X_f}\partial_T{A}\,
    \partial_X{A} \, dT dX
    \\&\quad\quad\quad+
     \int\limits_0^{T_f}\int\limits_{X_0}^{X_f} L_{\text{TW}}(\partial_X A,\partial_X Z,-\partial_T Z) \, dT dX\,,
\end{align*}
where  
\begin{align}
L_{\text{TW}}(V,K,\Omega)=\calL\big(\calW_*(V,K,\Omega), V, K, \Omega\big),
\end{align}
 is the action function for lattice traveling  waves
 and $\calW_*$  stands for a solution to \eqref{conopt}. The variations with respect to $A$ and $Z$ provide the two equations
\begin{align}
 \label{Eqn:Whitham0}
\begin{array}{cccccc}
\partial_T V &-&\partial_X\partial_V L_{\text{TW}}&=&0,\\
\partial_T \partial_\Omega L_{\text{TW}} &-&\partial_X\partial_K  L_{\text{TW}}&=&0,
\end{array}\,
\end{align}
which provide in combination with the integrability 
(the so-called ``conservation of waves'') condition 
\begin{align}
 \label{Eqn:WhithamInt}
\partial_T K+\partial_X \Omega=0,
\end{align}
 the three modulation equations with respect to the variational parameter set $V$, $K$, and $\Omega$. Setting $S=\partial _\Omega L_{\text{TW}}$ and regarding the Legendre transform
 \begin{equation}
 \label{Eqn:Whitham.LT}
 E_{\text{TW}}=S\, \Omega -   L_{\text{TW}},
 \end{equation}
as a function in $V$, $S$, and $\Omega$, the macroscopic parameter dynamics can equivalently be written as
 \begin{align}
 \label{Eqn:Whitham}
\begin{array}{cccccc}
\partial_T V &+&\partial_X\partial_V E&=&0,\\
\partial_T K &+&\partial_X\partial_S E&=&0,\\
\partial_T S &+&\partial_X\partial_K E&=&0\,.
\end{array}
\end{align}
This is a system of three nonlinear conservation laws
in which the densities completely determine the fluxes. The corresponding constitutive relations are all encoded in the equation of state \eqref{Eqn:Whitham.LT}, which depends via
\begin{align}
 \label{Eqn:Whitham.EOS}
E_{\text{TW}}(V,K,S)=\mathcal{E}\big(\mathcal{W}_*(V,K,S),V,K\big),
\end{align}
on the three-dimensional family of lattice waves, where $\calW_*$
now denotes a traveling wave with parameters $V$, $K$, and $S$, that is a critical point of the energy functional \eqref{Eqn:EFunctional} subject to the constraint $\calS(\calW_*,K)=S$ as in \eqref{Eqn:SFunctional}. By means of the reformulation
\eqref{Eqn:Whitham},  one readily verifies that 
the modulation equations \eqref{Eqn:Whitham} are of Hamiltonian type and imply (at least for smooth solutions) a fourth conservation law, namely
\begin{align} 
 \label{Eqn:Whitham.Energy}
\partial _T E_{\text{TW}}+\partial_X \big(\tfrac12\, \big(\partial_V E_{\text{TW}}\big)^2+ \partial_K E_{\text{TW}}\,\partial_S E_{\text{TW}}\big)=0\,.
\end{align}
The Whitham approach, therefore, predicts four macroscopic conservation  laws to be valid within each DSW  with only three of them being independent.  The first equation in \eqref{Eqn:Whitham} and the extension PDE \eqref{Eqn:Whitham.Energy} correspond to the two equations in \eqref{mde} and, therefore, possess an immediate microscopic interpretation. The second equation  in \eqref{Eqn:Whitham}
is just the consistency relation \eqref{Eqn:WhithamInt} between the wave number $K$ and the frequency $\Omega=\partial_S E_{\text{TW}}$, but due to the nonlinear nature of the underlying  lattice waves, it is very hard to relate these quantities to local space-time averages of microscopic observables. For the conservation law with density $S$, we likewise miss an interpretation on  the particle scale. For PDE models, this equation links to the conserved quantity  that stems from the shift invariance according to the Noether theorem, but its microscopic meaning remains unclear in the context of lattices and other particle systems. This is arguably even more
so in light of our earlier discussion about the
absence of translational invariance in the latter
context.

Despite the elegance of Whitham's variational approach to modulation theory, very little is known about the analytical properties of $E_{\text{TW}}$ and its derivatives. 
Except for integrable or very degenerate cases, it is not even clear whether \eqref{Eqn:Whitham} is really a hyperbolic PDE and how to characterize or compute the self-similar rarefaction waves that correspond to the DSW. This lack of qualitative (and even more so, 
quantitative) information was one of the main motivations for  following the alternative data-driven approach described in the main text.  

A particular
difficulty in the nonintegrable lattice setting is the dependence
of the wave profile on an advance-delay differential equation. 
This, in turn, appears to make explicit predictions intractable. Even the numerical simulation of the modulation equations can be quite cumbersome,
as discussed in \cite{DH08}. While the analysis of the modulation equations from this perspective
is certainly interesting and important, it is, as
discussed above, beyond the scope of the present article.

\bibliographystyle{unsrt}
\bibliography{Chong}

\end{document}